\documentclass[12pt]{article}
\usepackage{graphicx}
\usepackage{subcaption}
\usepackage[table]{xcolor}
\usepackage{eqnarray,amsmath}
\usepackage{makecell}
\usepackage{authblk}
\usepackage{setspace}

\definecolor{maroon}{rgb}{0.5, 0, 0}

\newcommand{\ie}{{\it i.e.}\ }
\newcommand{\eg}{{\it e.g.}\ }
\newcommand{\etal}{{\it et al.}\ }

\newcommand{\be}{\begin{eqnarray}}
\newcommand{\ee}{\end{eqnarray}}

\newcommand{\bes}{\begin{eqnarray*}}
\newcommand{\ees}{\end{eqnarray*}}
\newcommand{\ds}{\displaystyle}

\newcommand{\dd} \partial

\newcommand{\non} \nonumber

\newcommand{\ra}{\rightarrow}

\newcommand{\csin}{\bar{c}_{\rm s,\text{init}}}
\newcommand{\ueq}{U_{\rm eq}}

\newcommand{\G}{\mathcal{G}}
\newcommand{\Gh}{\hat{\mathcal{G}}}
\newcommand{\C}{\mathcal{C}}
\newcommand{\Q}{\mathcal{Q}}
\newcommand{\LL}{\mathcal{L}}

\newcommand{\N}{\mathcal{N}}
\newcommand{\PP}{\mathcal{P}}

\newcommand{\csm}{c_s^{\rm max}}
\newcommand{\Fms}{\mathcal{F}_{-}}
\newcommand{\Fm}{\mathcal{F}_{-}}
\newcommand{\Fmz}{\mathcal{F}_{-,0}}

\newcommand{\Fp}{\mathcal{F}_{+}}

\usepackage[margin=1in]{geometry}

\title{Generalised single particle models for high-rate operation of graded lithium-ion electrodes: systematic derivation and validation}
\author[1,2,5]{G. Richardson}
\author[1,2,6]{I. Korotkin}
\author[3]{R. Ranom}
\author[4]{M. Castle}
\author[2,4,7]{J. M. Foster}
\affil[1]{Mathematical Sciences, University of Southampton, University Rd., SO17 1BJ, UK}
\affil[2]{The Faraday Institution, Quad One, Becquerel Avenue, Harwell Campus, Didcot, OX11 0RA, UK}
\affil[3]{Faculty of Electrical Engineering, Universiti Teknikal Malaysia Melaka, 76100 Melaka, Malaysia}
\affil[4]{School and Mathematics and Physics, University of Portsmouth, Lion Terrace, PO1 3HF, UK}
\affil[5]{\tt g.richardson@soton.ac.uk}
\affil[6]{\tt i.korotkin@soton.ac.uk}
\affil[7]{\tt jamie.michael.foster@gmail.com}
\begin{document}

\maketitle

\begin{abstract}
{A derivation of the single particle model (SPM) is made from a porous electrode theory model (or Newman model) of half-cell (dis)charge for an electrode composed of uniformly sized spherical electrode particles of a single chemistry. The derivation uses a formal asymptotic method based on the disparity between the size of the thermal voltage and that of the characteristic change in overpotential that occurs during (de)lithiation. Comparison is made between solutions to the SPM and to the porous electrode theory (PET) model for NMC, graphite and LFP. These are used to identify regimes where the SPM gives accurate predictions. For most chemistries, even at moderate (dis)charge rates, there are appreciable discrepancies between the PET model and the SPM which can be attributed to spatial non-uniformities in the electrolyte. This motivates us to calculate a correction term to the SPM. Once this has been incorporated into the model its accuracy is significantly improved. Generalised versions of the SPM, that can describe graded electrodes containing multiple electrode particle sizes (or chemistries), are also derived. The results of the generalised SPM, with the correction term, compare favourably to the full PET model where the active electrode material is either NMC or graphite.}
\end{abstract}

\section{Introduction}
Lithium-ion batteries (LIBs) provide rechargeable energy storage at an unrivalled energy and power density, with a high cell voltage, and a slow loss of charge when not in use \cite{Blo17}. These characteristics mean that they are already ubiquitous in the consumer electronics sector and are being increasingly adopted for use in electric vehicles and off grid storage. However, particularly for vehicular propulsion, there are still major hurdles to be overcome in terms of lengthening service life, facilitating higher (dis)charge rates, and improving safety, particularly in high-rate regimes \cite{Vet05,Wan12}. Driven largely by the incumbent legislation to ban the combustion engine across large parts of the world before 2040, it has been predicted that the demand for LIBs will balloon from 45GWh/year (in 2015) to one of 390GWh/year in 2030 \cite{Zub18} and so improvements in LIB performance are needed as a matter of urgency.

A LIB pack is comprised of a collection of single electrochemical cells. Within each of these cells are three main components which facilitate the electrochemical reactions driving the electrical current; they are, two electrodes and the electrolyte. Both the positive and negative electrodes are themselves composites, being comprised of a porous network of microscopic electrode particles and a conductive polymer binder. The voids within this solid scaffold are filled with liquid electrolyte. Lithium (Li) can be inserted into and extracted from the particulate electrode material (intercalated and deintercalated). During discharge, the Li is extracted from the negative electrode material forming a free electron and a Li$^+$ ion. The ion is transported through the electrolyte (and separator diaphragm), and inserted into the positive electrode material. This ionic current is compensated by a flow of free electrons through the external circuit providing the useful electrical current. The charging process occurs similarly but with the ionic and electronic currents flowing in the opposite directions.

{Modelling of LIB performance takes place over a wide range of length scales ranging from atomistic scale simulations of battery materials to full device simulations and is reviewed by Franco in \cite{franco13}. The current work is focussed on electrode scale modelling of LIBs and is based on the approach developed by Newman and his co-workers in the mid 90's and early 2000's  \cite{Doy93,doyle96,Ful94,thomas02} which is often referred to as porous electrode theory (PET). The form of these models has since been justified using asymptotic homogenisation techniques in \cite{Ric12,Sch17}. In these models, partial differential equations for the Li concentration and electric potentials are solved across the electrode. In order that source terms (which capture the (de)intercalation reactions) can be accurately evaluated, at each point in the macroscopic dimension (\ie across the electrode) a microscopic problem is solved for the transport within the electrode particles.
In the Newman group's original work the electrode is assumed to have a one-dimensional slab-like geometry and the electrode particles are assumed spherically symmetric. Hence, since both the macro- and microscopic problems are effectively one-dimensional, their work is often referred to as a pseudo-two-dimensional approach \cite{Jok16}. The original PET works also choose to model lithium transport within the electrode particles via a simple linear diffusion model. With recent improvements in the understanding of the chemistry of these active (electrode) materials it has become clear that a linear diffusion model does not provide a good description of this transport process. Hence many recent works have focussed on incorporating more realistic lithium transport models in the active materials into PET. These range from nonlinear diffusion models calibrated against experimental data \cite{Ecker2015a,Ecker2015b} to Cahn-Hilliard models for phase change materials \cite{cogswell13,dargaville13,ferguson14}. Here we will employ the former approach and restrict our attention to nonlinear diffusion models of lithium transport in the active materials. The original PET formulation has also been generalised by posing the problem, on both  macro- and micro-scales, in higher numbers of dimensions. This enables the effects of electrode particle shape to be investigated, and also spatially inhomogeneities in electrode discharge that could arise, for example, from non-uniform heating or the positioning of the electrode tabs, see \cite{Fos15,Kos18}.}


Whilst it is difficult to overstate the success and utility of the PET approach, it can be criticised for being relatively expensive to solve. This is a particularly significant difficulty when it is being used as a tool to optimise cell design \cite{cheng19}, or {extended to model} to 2- or 3-macroscopic dimensions \cite{Yi13,Dan16}, or as a tool in parameter estimation studies \cite{bizeray18,sethurajan19}. Its multiscale nature means that the underlying equations are posed over two separate spatial dimensions and, since these equations are nonlinear, obtaining solutions is a task that needs to be tackled numerically. This has motivated many authors to consider simplified versions of the PET. Perhaps the most well-known model of this type is the single particle model, or SPM, which results from assuming that each of the electrode particles are equally sized spheres, and then arguing that the electrochemical reactions occur roughly uniformly throughout the electrodes so that the active material in the electrode particles (de)lithitates at the same rate independent position in the electrode \cite{Mou16,Rah13}. In this way, finding model solutions is reduced to the task of solving a single spherical transport problem inside a `representative' particle in each electrode. In this context we note that the thesis of Ranom \cite{ranom15}, from which the current work stems, contains a systematic derivation of the leading order SPM from the PET model. 

{After completion of this work, we became aware of another article in-progress that employs asymptotic methods to simplify the PET model \cite{marquis19}. In \cite{marquis19}, the asymptotic limit of large electrode and electrolyte conductivities and large electrolyte diffusivity is taken; this is a different (and in fact complementary) limit to that taken here. Their limit recovers a variant of the SPM at leading order because the gradients in the electrolyte concentration, electrolyte potential, and  electrode potentials are all small and this gives rise to  homogeneous behaviour of the electrode particles and hence leads to the SPM. In contrast, in our limit the variation in the overpotential across the electrode is small in comparison the typical variation in the equilibrium potential as lithium is removed/inserted into the electrode and it is this that leads to the SPM. Whilst both limits recover an SPM at leading order, the correction terms are different. If the model parameters for a particular case are not appropriate for the limit considered here, we encourage the reader to also consider the reduced model in \cite{marquis19}. We also note the work of Moyles \etal \cite{moyles18} which also derives an asymptotic reduction of a PET model, but does so by first volume averaging lithium transport over the electrode particles in the PET model, and thereby implicitly assumes transport within individual electrode particles is rapid.}

In order to properly understand for which electrode chemistries it is appropriate to use the SPM approximation we will, throughout this work, restrict our attention to half-cell configurations, noting that the extension to full cells is {straightforward}. Such half-cell configurations are comprised of a single porous electrode (either a cathode or anode) being (dis)charged, through a separator, against a metallic Li counter electrode {\cite{New93,Wu00}}, see figure {\ref{pretty}}. {We will allow the electrodes within our half-cells to contain more than one size of electrode particle and/or more than one chemistry, and this will allow us to derive generalised versions of the SPM applicable to graded electrodes.} Motivated by the need to provide a useful tool for the practitioner we validate our approximations, as far as possible, against realistic data sets for the PET model. Here these data sets are primarily based upon two works from the Ecker group \cite{Ecker2015a,Ecker2015b} which adopt a combined experimental and theoretical approach validating the results of their PET model simulations against data collected from real cells.

{In the next section of the paper we describe the PET model as well as the boundary and initial conditions with which it should be supplemented in order to mimic a half-cell configuration. Then, in \S\ref{sec:res} we describe} {a generalisation of the standard SPM that is applicable to graded electrodes and ones with multiple electrode particle chemistries. We also outline how higher order terms, which capture the effects of spatial and temporal variations in the electrolyte properties, can be incorporated into this generalised SPM. Including these higher order terms requires that a one-dimensional system of PDEs be solved in the electrolyte, in addition to the one-dimensional PDEs in the electrode particles. However introducing this extra complexity into the simplified model significantly enhances its accuracy as we demonstrate with the aid of some examples: (i) electrodes with a uniform particle size and (ii) graded electrodes with two sizes particle segregated into separate regions.} {In \S\ref{sec:nondim} we introduce scalings and rewrite the PET in dimensionless form. This nondimensionalisation facilitates the asymptotic analysis which is the subject of \S\ref{asy}. Here, the SPM, as well as its generalisised version for graded electrodes, and its higher order correction terms, is derived systematically from the PET. Finally, in \S\ref{conc} we draw our conclusions.}


\section{\label{probform}Problem formulation}
\begin{figure}          \centering
\includegraphics[width=0.65\textwidth]{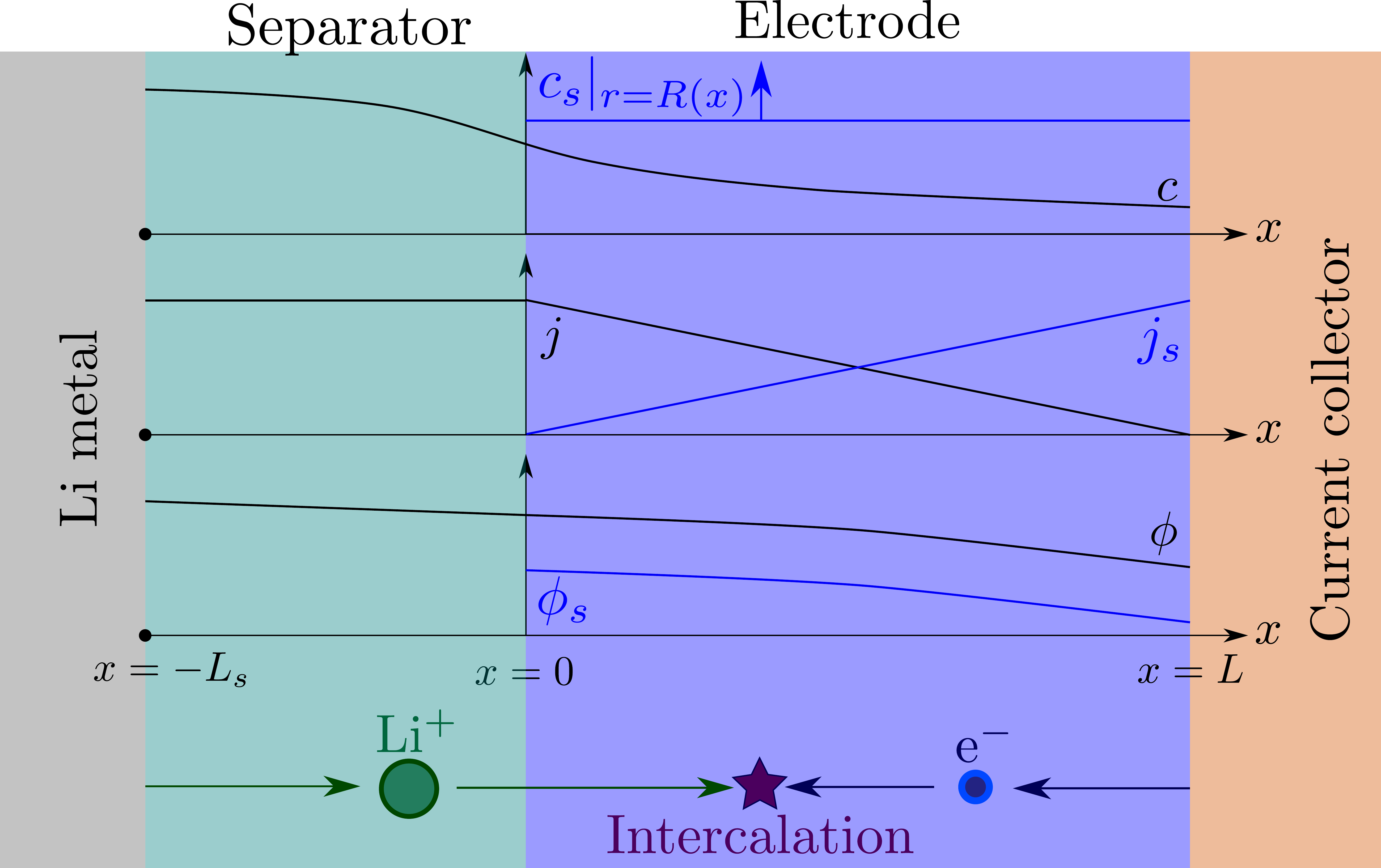}
\caption{A schematic of the half cell geometry as well as the independent variables and snapshot of their qualitative profiles during charging if the electrode is an anode, or discharging if the electrode is a cathode.}
\label{pretty}
\end{figure}

Here, we consider a PET model posed on the model geometry shown in figure \ref{pretty}. The equations governing ionic transport through the electrolyte are\cite{ciucci,Ful94,ranom15,Ric12}
\be \label{dim1}
\epsilon_l(x)\frac{\partial c}{\partial t}+\frac{\partial \Fms}{\partial x}=0,\quad 
\Fms=-\mathcal{B}(x)D(c)\frac{\partial c}{\partial x}-(1-t^+)\frac{j}{F} \quad \mbox{in} \quad -L_s < x<L.
\ee
where, $x$ and $t$ denote position through the electrode and time respectively, $\epsilon_l$ is the local volume fraction of electrolyte, $c$ is the molar concentration of ions (the Li and counter ion concentrations are equal throughout the bulk owing to the extremely short Debye length of the solution) in the electrolyte, $\Fms$ is the effective flux of anions across the electrolyte, $D$ is the ionic diffusivity of the electrolyte, $\mathcal{B}$ is the permeability factor (often estimated using the ad-hoc relation offered by Bruggeman which takes ${\mathcal B}=\epsilon_l^{1.5}$ \cite{Bru35}), $t^+$ is the transference number, $j$ is the ionic current density and $F$ is Faraday's constant. In contrast to some other authors, we opt to write the conservation equation in terms of the anion flux, $\Fms$, rather than the cation flux, {$\Fp$}. We make this choice, because the anion is not (de)intercalated into the {electrode particles} and as a consequence its governing equation takes a conservative form which is more susceptible to the numerical treatment that we employ later in this work. This would not be the case if the conservation equation is written for the lithium cation which must contain a source/sink terms to account for the (de)intercalation process. The ionic current in the electrolyte obeys
\be \label{dim3}
\frac{\partial j}{\partial x}=F b(x) G, \quad 
j=-\mathcal{B}(x)\kappa(c)\left(\frac{\partial \phi}{\partial x}-\frac{2RT}{F}\frac{1-t^+}{c}\frac{\partial c}{\partial x}\right)\quad \mbox{in} \quad -L_s < x<L.
\ee
Here $b$ is the Brunauer-Emmett-Teller (BET) surface area (with units of $1/m$), $G$ is the reaction rate {which is zero in the separator and given by the Butler-Volmer equations in the electrode (where it is the flux per unit area of Li through the surface of the electrode particles)}, $\kappa$ is the ionic conductivity, $\phi$ is the electric potential in the electrolyte, $R$ is the (molar/universal/ideal) gas constant and $T$ is absolute temperature. The electric transport through the solid scaffold is governed by
\be \label{dim5}
\frac{\partial j_s}{\partial x}=-F b(x) G,\quad 
j_s=-{\sigma}\frac{\partial \phi_s}{\partial x} \quad \mbox{in} \quad 0< x<L
\ee
where $\phi_s$, $j_s$ and $\sigma$ are the electric potential, current density and conductivity of the electrode. We will refer to equations {(\ref{dim1})-(\ref{dim5})} as the macroscopic equations because their independent spatial variable, $x$, measures the macroscopic distance across the thickness of the electrode (cf. figure \ref{pretty}).

\paragraph{Boundary conditions}
We will assume that the half-cell (dis)charges according to some specified current {supply/demand}, $I(t)$, and as such boundary conditions to supplement the sixth order system of PDEs posed on the macroscale, {(\ref{dim1})-(\ref{dim5})}, require: (i) That the Li-metal supplies an ionic current density of $I(t)/A$ to the electrolyte in the separator (on $x=-L_s$). We further require, (ii), that at the current collector ($x=L$) there is injection of current of density $I(t)/A$ into the solid phase. In addition, (iii), no current flows from the solid phase into the separator at $x=0$. A reference potential in the electrolyte is provided at the edge of the separator where it meets the Li-metal ($x=-L_s$). Finally, (v) and (vi), we specify that there should {be} no anion flux in the electrolyte at the interfaces where it meets both the separator and current collector. In summary we have
\be\label{chbcs1dim}
\label{chbcs2dim}  j|_{x=-L_s} = \frac{I(t)}{A}, \quad \Fms|_{x=-L_s} = 0, \quad \phi|_{x=-L_s} =0,\\
\label{shortname} j_s|_{x=0} = 0,  \quad j_s|_{x=L} = \frac{I(t)}{A}, \quad \Fms|_{x=L} = 0.
\ee

\paragraph{The Butler-Volmer reaction rate}
{It remains to specify the reaction rate $G$. In the separator ($-L_s < x<0$), where there are no electrode particles, this is zero. In the electrode ($0 \leq x< L$) the reaction rate is determined by the Butler-Volmer (BV) equation \cite{Ful94,NewmanBook,New75} such that $G$ is given by
\be
\label{Dim-BV}
G=\left\{\begin{array}{ccc}
0 & \mbox{in} & -L_s < x<0, \\
\ds 2 k c^{1/2} \left( c_{s}\rvert _{r=R(x)} \right)^{1/2} \left( \csm -c_{s}\rvert _{r=R(x)}\right)^{1/2} \sinh\left(\frac{F\eta}{2RT}\right) &  \mbox{in} & 0 \leq x< L,  \end{array} \right.
\ee}
where the overpotential is given by
\be
\label{Dim-OP}
\eta=\phi_s-\phi-\ueq(c_s|_{r=R(x)}).
\ee
Here, $\ueq$, $\csm$ and $R(x)$ are the equilibrium potential, the maximum concentration of lithium that can be stored in the electrode material, and the radii of electrode particles. 

\paragraph{Equations on the microstructure}
Evaluating the (de)intercalation rates, $G$, necessitates solving for the solid-state lithium concentration on the surfaces of the electrode particles throughout the electrode. Whilst there is considerable debate about the correct solid state transport model, particularly for phases separating electrode materials \cite{Bai11,Li14,Per10}, 
here, we opt to solve non-linear diffusion equations in the active material particles and as such we have
\be
\label{dim7}
\frac{\partial c_{s}}{\partial t}=\frac{1}{r^{2}}\frac{\partial }{\partial r}\left( r^{2} D_{s} (c_{s})\frac{\partial c_{s}}{\partial r}\right) \quad \mbox{in} \quad 0< r<R(x)\ \  \mbox{and} \ \  0< x<L.
\ee
Here $D_s$ and $r$ are the concentration-dependent diffusivity and radial coordinate within the particles respectively. Equations (\ref{dim7}) are closed by supplying them with the boundary conditions
\begin{align}
\label{Dim-BC 4}
c_s \,\, \mbox{bounded} \,\, \mbox{on} \,\, r=0, \qquad -D_{s}(c_{s}) \frac{\partial c_{s}}{\partial r}\bigg\rvert_{r= R(x)}=G.
\end{align}
The former is a regularity conditions required to eliminate singular behaviour at the origin whilst the latter ensures that there is the requistite Li flux across the surface owing to the reactions taking place there. We will refer to the (\ref{dim7})-(\ref{Dim-BC 4}) as the microscopic transport problem.

\paragraph{The half-cell potential} 
The half-cell potential, $V$, is given by the expression
\be \label{zap}
V(t)= \phi_s\big\rvert_{x=L}.
\ee
Note that it is straightforward to include an Ohmic drop across the current collector interface caused by a contact resistance ${\cal R}$ by replacing this expression by one of the form $V(t)= \phi_s\big\rvert_{x=L}-{\cal R} I$.

\paragraph{Initial conditions}
The parabolic equations (\ref{dim1}) and (\ref{dim7}) require initial conditions on the electrolyte ion concentration $c$ throughout the electrolyte as well as $c_s$, the concentration of Li in the electrode particles. We assume that the electrode is initially in equilibrium so that both $I(0)=0$ and $G(x,0)=0$. As such we require a uniform salt concentration in the electrolyte
\be
c\big\rvert_{t=0}=c_{\text{init}} \quad \mbox{in} \quad  -L_s< x<L.
\ee
and a uniform lithium concentration throughout all the electrode particles
\begin{eqnarray}
c_{s}\big\rvert_{t=0}= c_{s,\text{init}}  \quad \mbox{in} \quad 0< r<R(x)\ \  \mbox{and} \ \  0< x<L.
\end{eqnarray}
It follows from these conditions, and the model equations, that initially $\phi(x,0)=0$ and $\phi_s(x,0)=\ueq(c_{s,\text{init}})$.

\paragraph{Electrode geometry} We assume that all electrode particles are spheres, with radii that vary on the relatively long lengthscales assoiated with the macroscopic thickness (but not on the much shorter lengthscale of individual electrode particles, i.e., the microscale). This gives rise to the following relationship between the BET surface area, particle radius, and volume fraction of electrode particles
\begin{align} \label{geoeqn}
\epsilon_s(x)=\frac{b(x)R(x)}{3},
\end{align}
where $\epsilon_{s}$ is the local volume fraction of electrode particles. The relationship (\ref{geoeqn}) follows from noting that the volume fraction of particles within a small representative elementary volume is the product of the volume of a single particle, namely $4 \pi R^{3}$, and the number of particles within the REV, namely $b / (4 \pi R^{2})$. The volume fraction, $\epsilon_s$, is related to the volume fraction of liquid in the electrode via
\be
\epsilon_l(x) = 1- \epsilon_{\text{inert}}(x) - \epsilon_{s}(x), \label{finaleqn}
\ee
where $\epsilon_{\text{inert}}$ is the volume fraction of electrochemically inert material (e.g., polymer binder or the conductivity enhancing carbon black).

\section{\label{sec:res}Results}
We begin by stating the generalisation of the single particle model (SPM) approximation to graded electrodes. This is based on the assumption that ${\cal U}$ the characteristic change in overpotential that occurs as lithium is intercalated into (or removed from) the electrode material is much larger than the thermal voltage {$V_T =RT/F \approx 26$mV}. Formally we require
\bes
\lambda \ll 1 \quad \mbox{where} \quad \lambda = \frac{F {\cal U}}{RT}.
\ees
It is thus applicable to materials such as graphite, NMC or LCO (though not to LFP where ${\cal U} \leq 26$mV). We then give a recipe for extending the SPM to incorporate further terms, which depend upon the electrolyte behaviour, and considerably enhance the accuracy of the method. Henceforth we will refer to the corrected SPM, which includes the additional terms, as the `corrected SPM'. In order to illustrate the method we consider two examples. The first of these (described in \S \ref{SPM}) is the single particle model which applies to electrodes comprised of uniform sized electrode particles (\ie with no grading) while the second (described in \S \ref{DPM}) is the double particle model which applies to electrodes which are graded with two particle sizes occupying different regions of the electrode (one near the separator and one near the current collector). In both cases we compare results of the SPM and corrected SPM to the full numerical solution to the PET model \eqref{dim1}-\eqref{zap} for electrodes of several different chemistries. Derivation of the approximate models, from the PET model, is deferred until \S \ref{asy}.

\begin{figure} \centering
\includegraphics[width=0.495\textwidth]{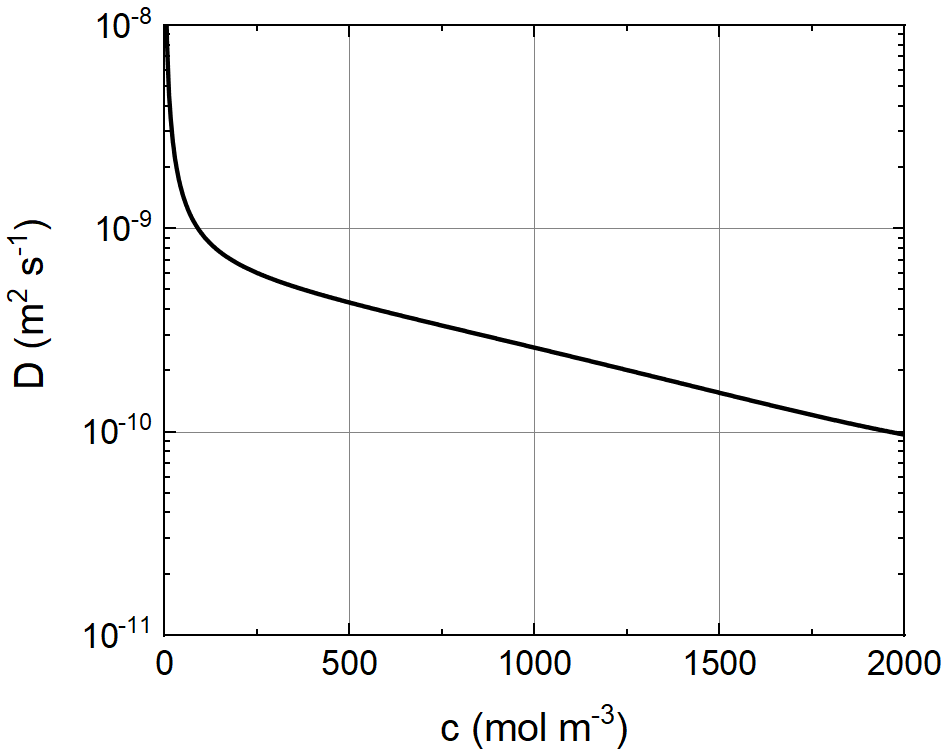}
\includegraphics[width=0.495\textwidth]{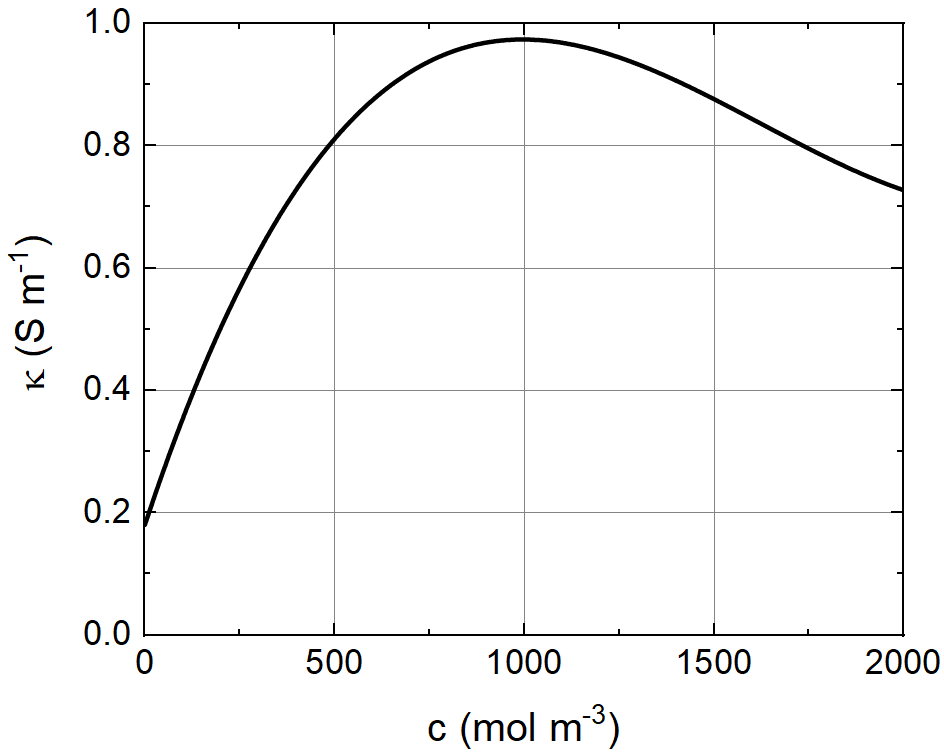}\\
\medskip
\includegraphics[width=0.495\textwidth]{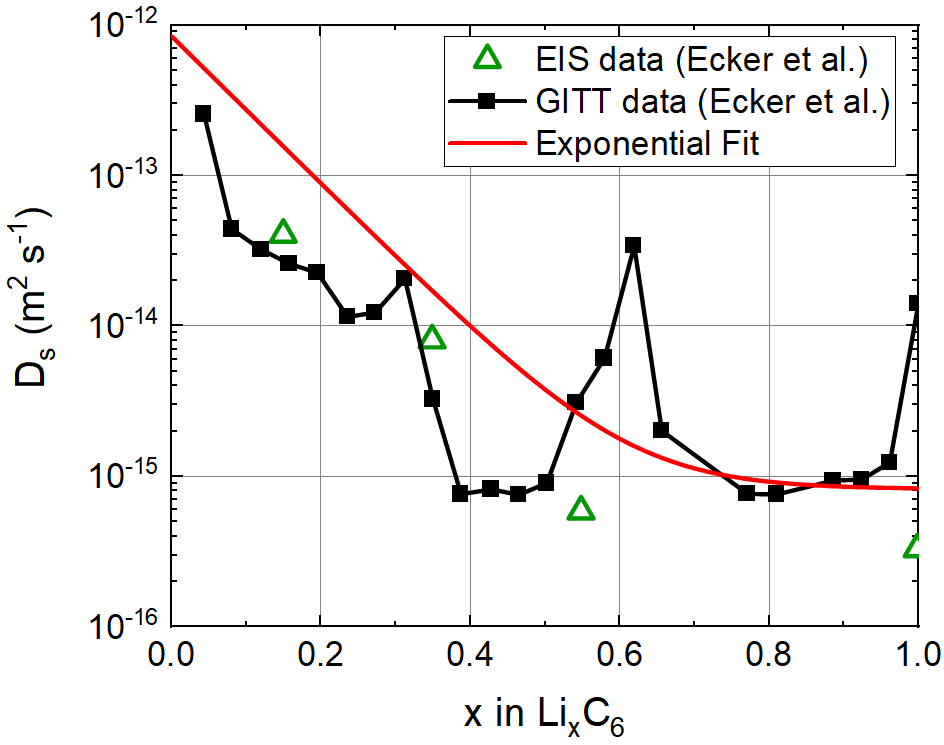}
\includegraphics[width=0.495\textwidth]{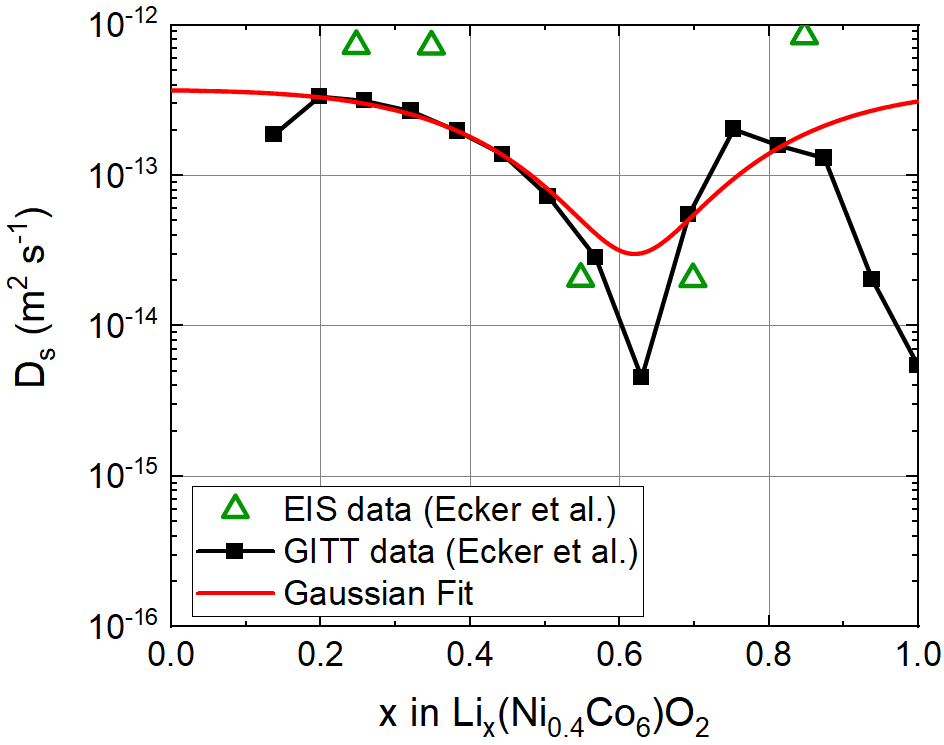}\\
\caption{Ionic diffusivity (top left) and conductivity (top right) of the electrolyte from \cite{Ecker2015a}, non-linear diffusivity of graphite ($\text{LiC}_6$) anode (bottom left) and $\text{Li}(\text{Ni}_{0.4}\text{Co}_{0.6})\text{O}_2$ cathode (bottom right). Experimental data from \cite{Ecker2015a} and approximate fits.}
\label{Dliq}
\end{figure}


\begin{figure} \centering
\includegraphics[width=0.495\textwidth]{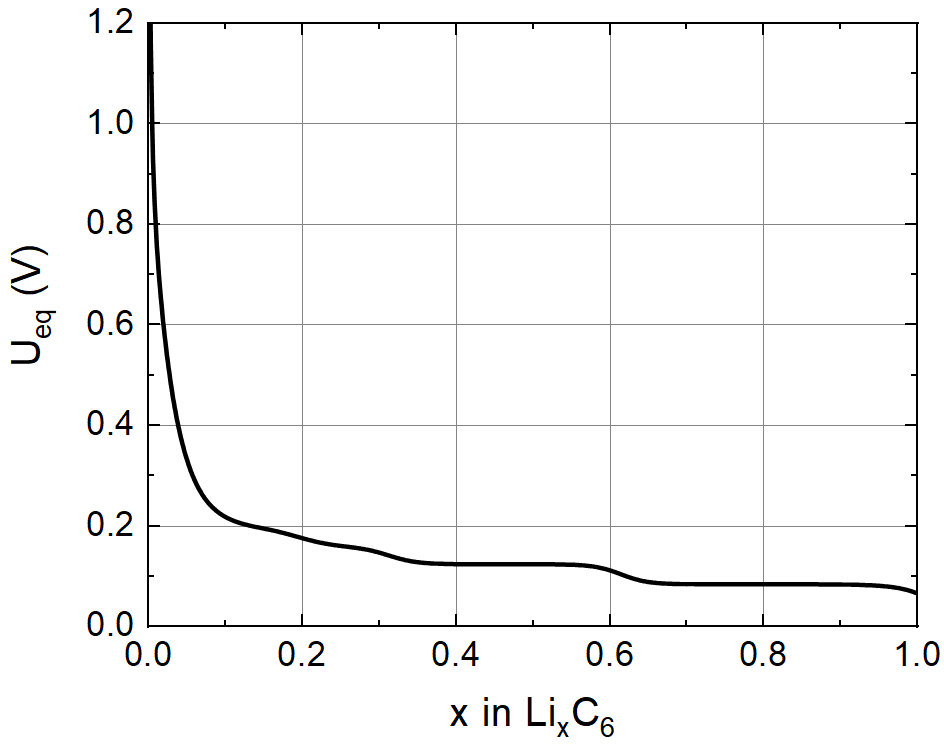}
\includegraphics[width=0.495\textwidth]{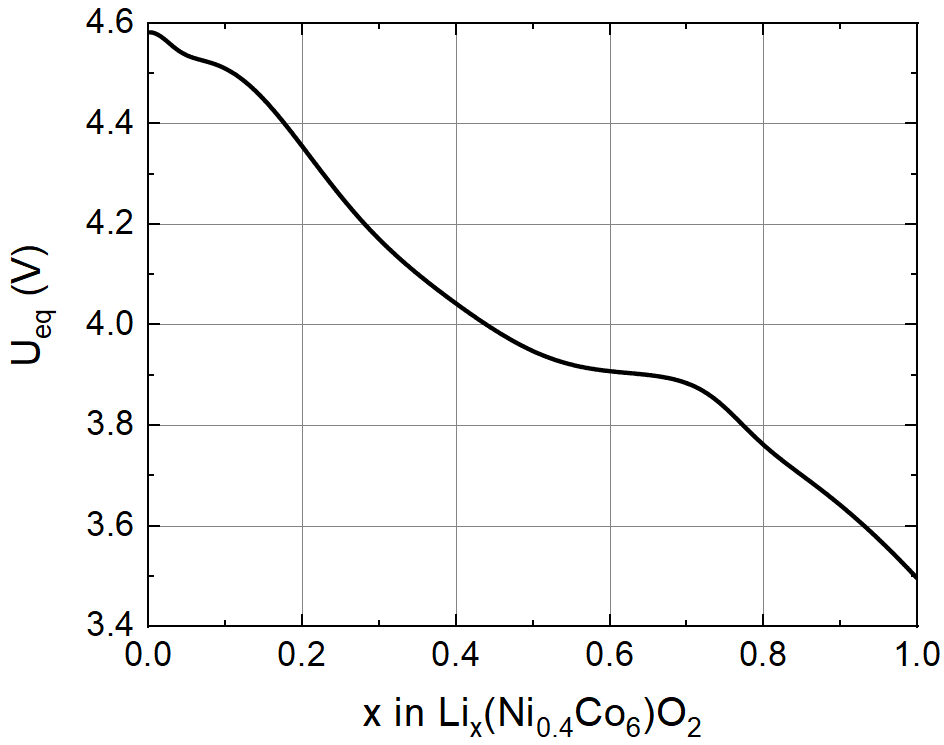}\\
\medskip
\includegraphics[width=0.495\textwidth]{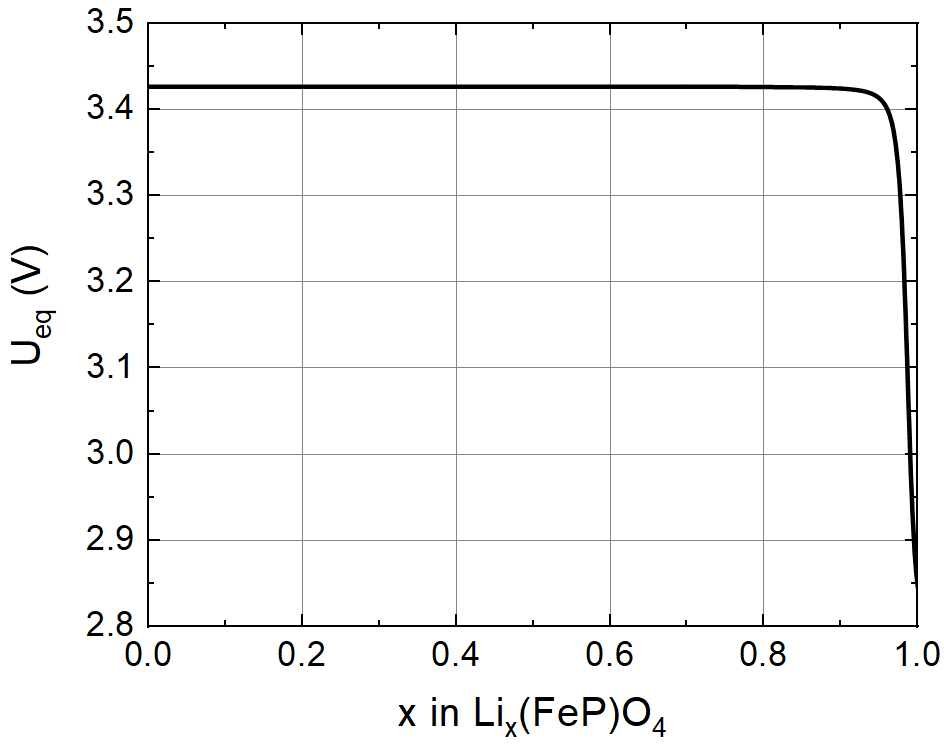}\\
\caption{Equilibrium potential of graphite ($\text{LiC}_6$) anode \cite{Ecker2015a} (top left), $\text{Li}(\text{Ni}_{0.4}\text{Co}_{0.6})\text{O}_2$ cathode \cite{Ecker2015a} (top right), and LFP ($\text{LiFePO}_4$) electrode \cite{ranom15} (bottom).}
\label{Ueq}
\end{figure}

\paragraph{The leading order approximation.}
The most basic approximation to the Newman model \eqref{dim1}-\eqref{zap} in the scenario described above is the following generalisation of the SPM:
\be
\frac{\partial c_{s}}{\partial t}=\frac{1}{r^{2}}\frac{\partial }{\partial r}\left( r^{2} D_{s} (c_{s})\frac{\partial c_{s}}{\partial r}\right) \quad \mbox{in} \quad 0 \leq r \leq R(x),~~~~ \label{appr1} \\
c_s \,\, \mbox{bounded} \,\, \mbox{on} \,\, r=0, \qquad c_s|_{r=R(x)}=\C(t), \label{appr2} \qquad c_s|_{t=0}=c_{s,\text{init}},\\
\int_0^L b(x) G(x,t) dx=\frac{I(t)}{A F} \quad \mbox{where} \quad G=\left. - D_s(c_s) \frac{\dd c_s}{\dd r} \right|_{r=R(x)} \label{appr3}
\ee
Here the function $\C(t)$ in \eqref{appr2} is chosen so that the integral constraint \eqref{appr3} is satisfied. The leading order approximation to the voltage, $V(t)$, of the half-cell is calculated from the solution of this problem \eqref{appr1}-\eqref{appr3} via the relation
\be
V(t) \approx \ueq(\C(t)). \label{vapprox_0}
\ee

\paragraph{A higher order approximation.} A more accurate expression for $V(t)$ can be calculated from the solution to \eqref{appr1}-\eqref{appr3} but additionally requires a one-dimensional problem for the electrolyte be solved. This expression reads
\be
V(t) \approx \ueq(\C(t))+ \frac{\displaystyle \int_0^L b(x) R(x)\left[ \eta(x,t) +\phi(x,t) - \int_x^L \frac{j_s(x',t)}{\sigma(x')} dx' \right] dx}{\displaystyle \int_0^L b(x) R(x) dx} . \label{vapprox_1}
\ee
Here $\phi(x,t)$, the potential in the electrolyte, and $\eta(x,t)$, the associated overpotential, are calculated from the solution to the one dimensional electrolyte problem
\be
\epsilon_l(x)\frac{\partial c}{\partial t}+\frac{\partial \Fms}{\partial x}=0,\quad 
\Fms=-\mathcal{B}(x)D(c)\frac{\partial c}{\partial x}-(1-t^+)\frac{j}{F},\label{lyte1} \\
\frac{\partial j}{\partial x}=F b(x) G(x,t), \quad 
j=-\mathcal{B}(x)\kappa(c)\left(\frac{\partial \phi}{\partial x}-\frac{2RT}{F}\frac{1-t^+}{c}\frac{\partial c}{\partial x}\right),
\label{lyte2}
\ee
with
\be
j|_{x=-L_s} = \frac{I(t)}{A}, \quad \Fms|_{x=-L_s} = 0, \quad \phi|_{x=-L_s} =0,\quad
 \quad \Fms|_{x=L} = 0,  \label{lyte3} \\
\text{and} \quad c|_{t=0} = c_{\text{init}},
 \ee
where 
\be
j_s(x,t)=\frac{I}{A} - j(x,t), \qquad \eta(x,t)=2 \frac{RT}{F} \mbox{arcsinh} \left(\frac{G(x,t)}{2 k [ (\csm - \C(t)) \C(t) c]^{1/2} }\right), \label{lyte4}
\ee
and $G(x,t)$ is obtained from the solution to the leading order problem \eqref{appr1}-\eqref{appr3}.
Notably the approximated expression for $V(t)$ calculated in \eqref{vapprox_1} is a formally accurate approximation to the PET model \eqref{dim1}-\eqref{zap}  (as explained in \S \ref{asy}), for all discharge rates, in the case that the electrode is composed of uniformly sized particles of a single material. However, even where this is not the case (\eg if the electrode is graded), it is still formally accurate provided that the (dis)charge rate is not excessively large in comparison to the characteristic timescale for transport within the electrode particles. More specifically, it is still formally accurate provided
\bes
{\cal Q} \ll 1 \quad \mbox{where} \quad 3 {\cal Q} = \frac{\mbox{timescale for Li diffusion into electrode particle}}{\mbox{timescale for cell discharge}}.
\ees  
The dimensionless parameter $\cal Q$ will be defined rigorously below in \S\ref{sec:nondim}. Since this requirement is usually satisfied, even at moderate to aggressive discharge rates, it usually offers a significant improvement over the leading order approximation \eqref{vapprox_0}.

\subsection{The Single Particle Model and comparison to the PET model \label{SPM}}
In the case where all the electrode particles are of uniform size $R$, across the width of the electrode the leading order equations \eqref{appr1}-\eqref{appr3} simplify considerably since $c_s=c_s(r,t)$ and $G=G(t)$ so that in this case we need solve only a single microscopic transport problem in $r$, namely \eqref{appr1} subject to the boundary conditions
\be
c_s \,\, \mbox{bounded on} \,\, r=0, \,\, \left. D_s(c_s) \frac{\dd c_s}{\dd r} \right|_{r=R}=-G(t) \quad \mbox{where} \quad G(t)=\frac{I(t)}{A F \int_0^L b(x) dx}, ~~~
\ee
from which we can then evaluate
\be
\C(t)=c_s|_{r=R}
\ee
and use the functions $G(t)$ and $\C(t)$ thus determined as inputs in the electrolyte model \eqref{lyte1}-\eqref{lyte4} which we can then solve to obtain the data necessary to compute the corrected voltage via \eqref{vapprox_1}.

\begin{figure} \centering
\includegraphics[width=0.495\textwidth]{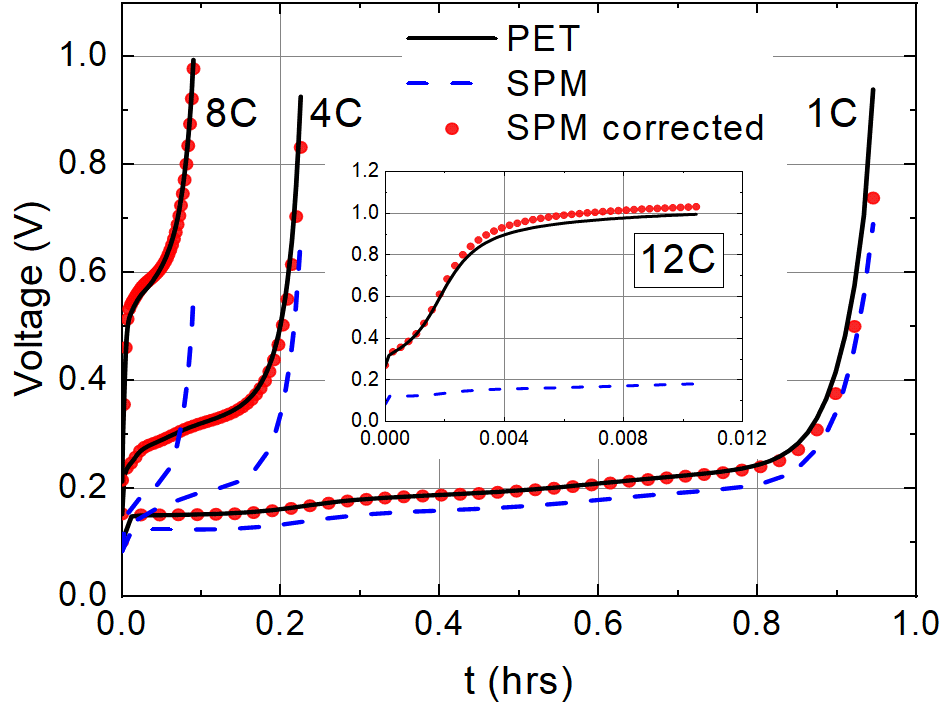}
\includegraphics[width=0.495\textwidth]{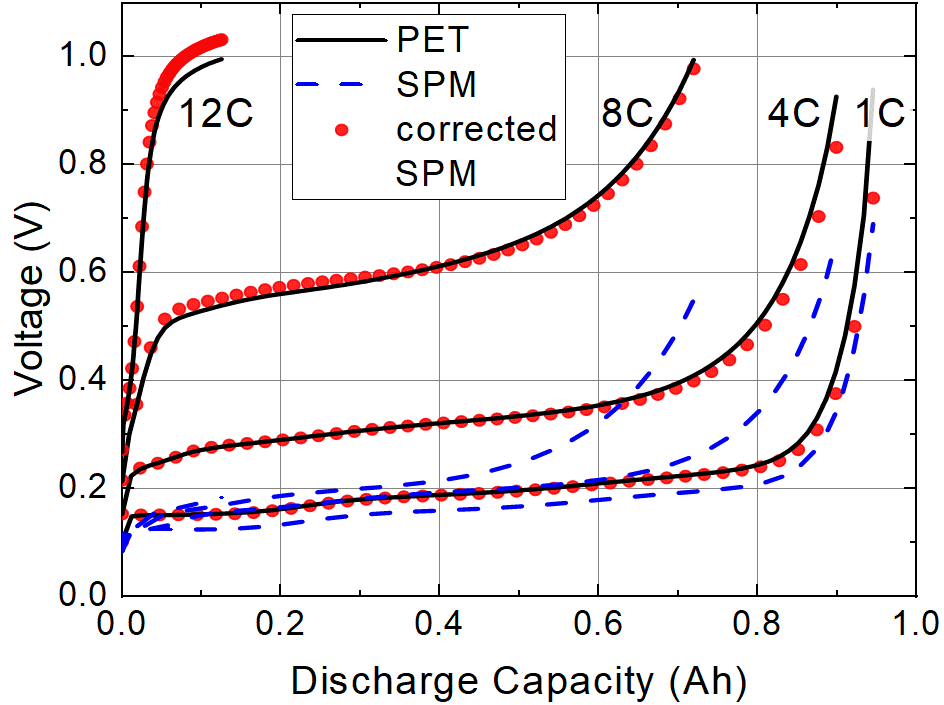}\\
\medskip
\includegraphics[width=0.495\textwidth]{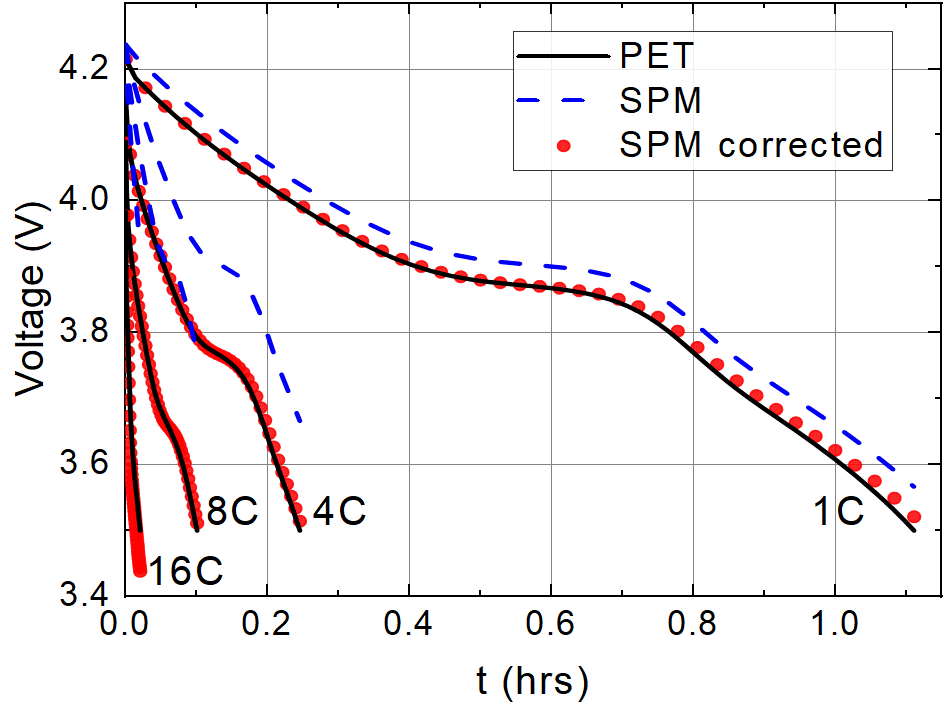}
\includegraphics[width=0.495\textwidth]{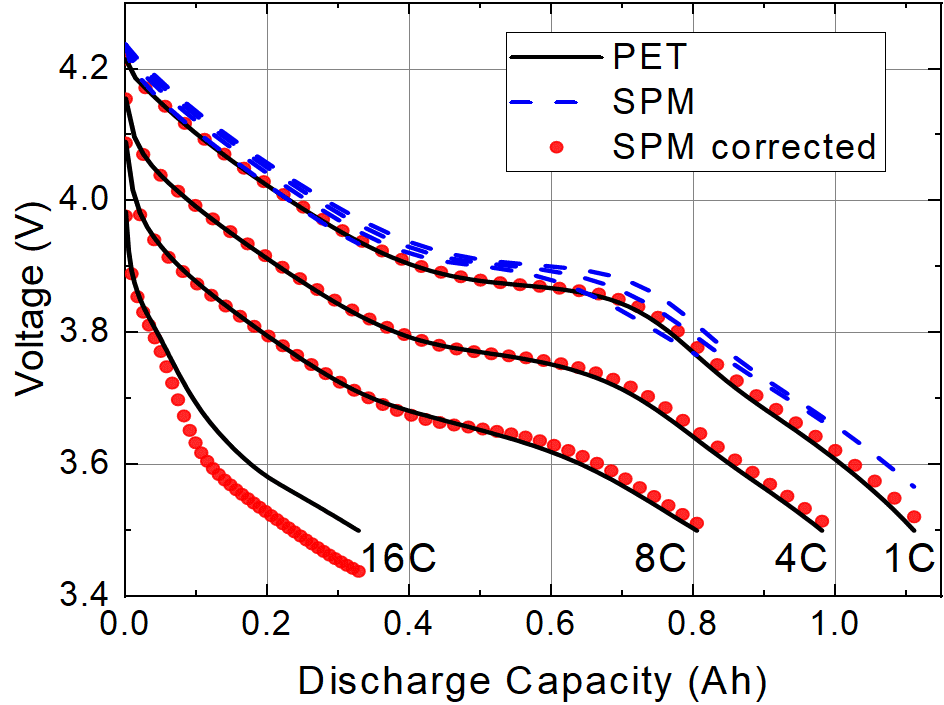}\\
\medskip
\includegraphics[width=0.495\textwidth]{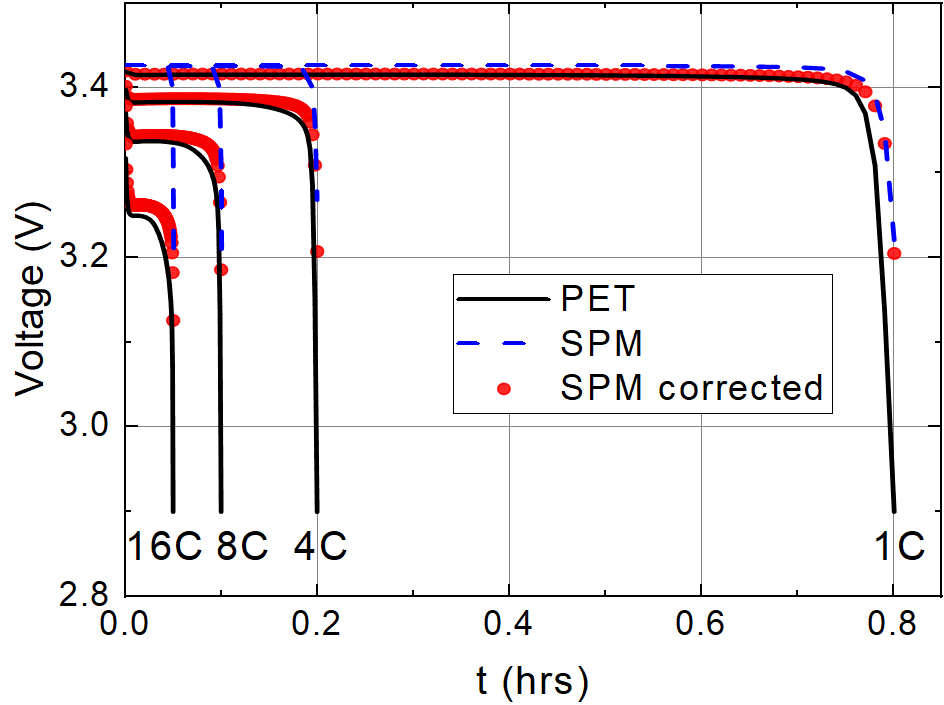}
\includegraphics[width=0.495\textwidth]{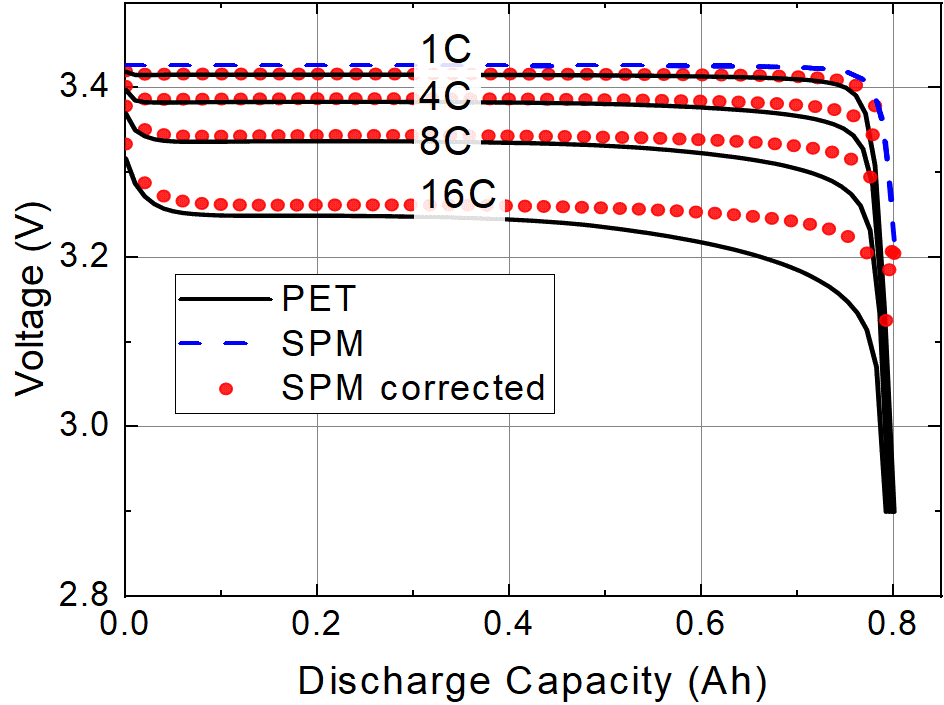}\\
\caption{Cell potentials $V$ calculated using the full Newman PET model, SPM and corrected SPM for graphite ($\text{LiC}_6$) anode (top row), $\text{Li}(\text{Ni}_{0.4}\text{Co}_{0.6})\text{O}_2$ cathode (middle row) and LFP ($\text{LiFePO}_4$) electrode (bottom row) at different discharge rates. {Results are shown against both time (left) and discharge capacity (right).}}
\label{SPM_PET_plot}
\end{figure}

\begin{figure} \centering
\includegraphics[width=0.495\textwidth]{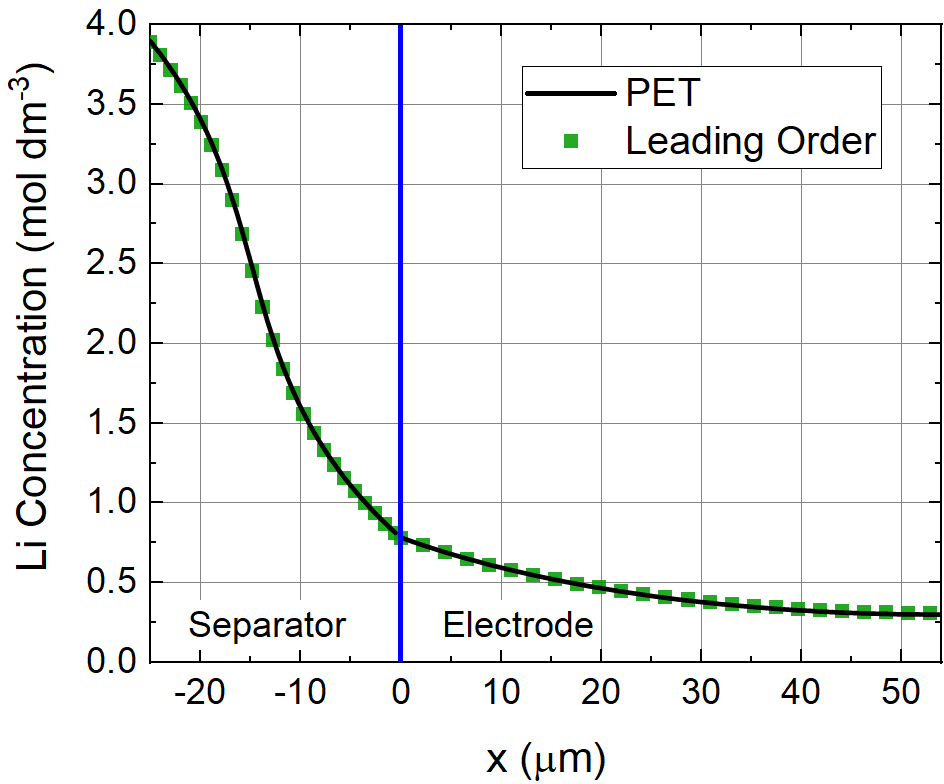}
\includegraphics[width=0.495\textwidth]{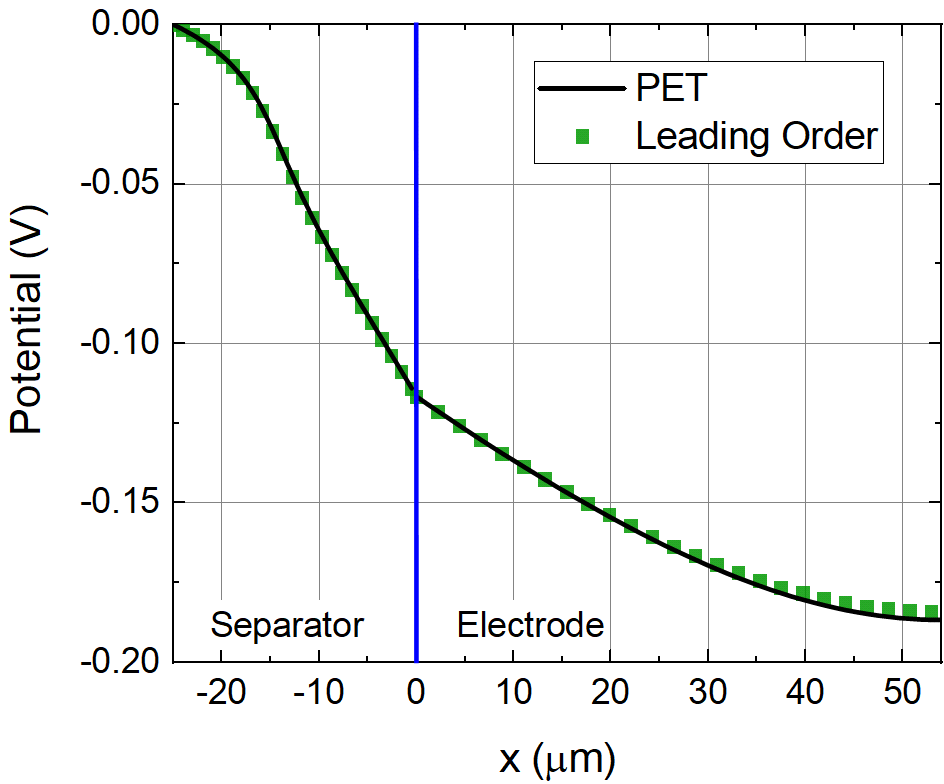}\\
\medskip
\includegraphics[width=0.495\textwidth]{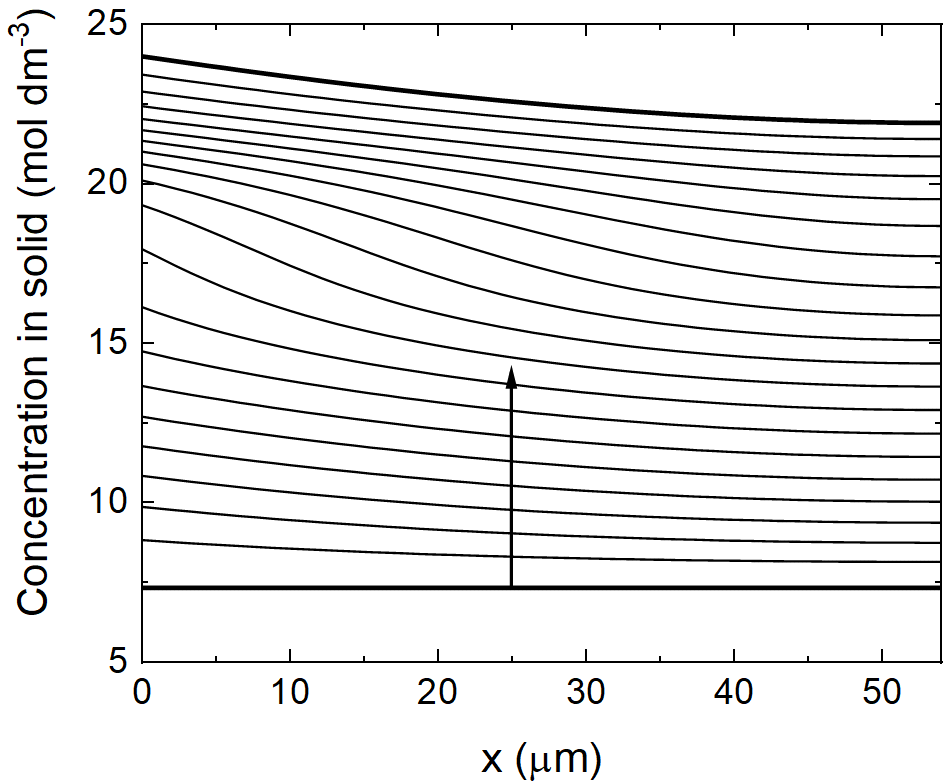}
\includegraphics[width=0.495\textwidth]{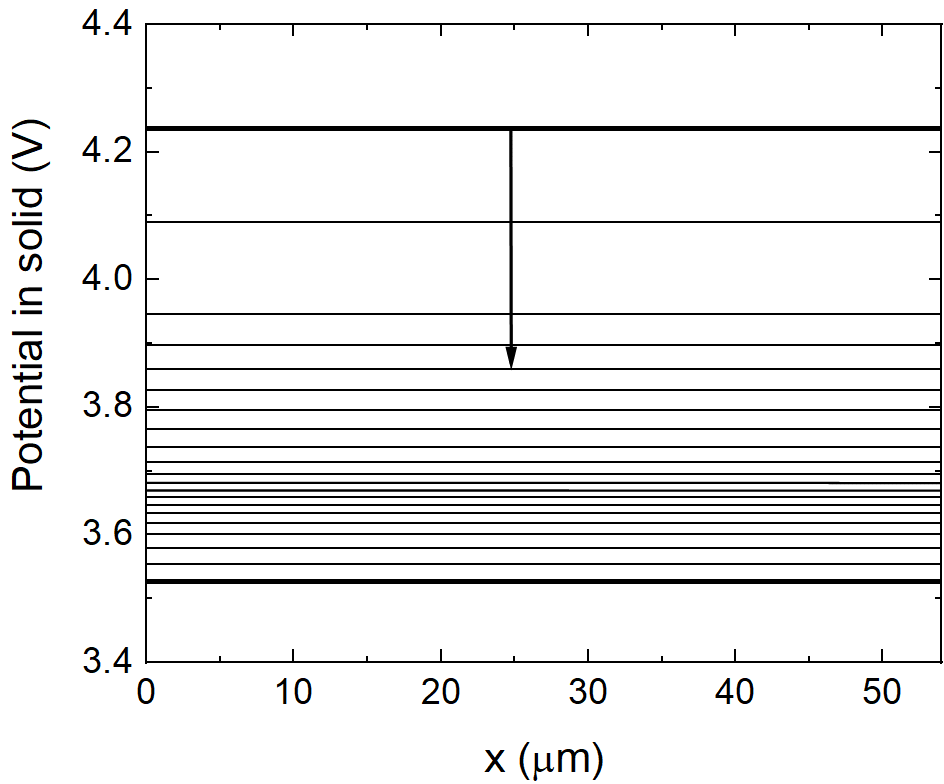}\\
\caption{Concentration of Li$^{+}$ ions (top left) and potential (top right) in the electrolyte {across a half-cell with a $\text{Li}(\text{Ni}_{0.4}\text{Co}_{0.6})\text{O}_2$ electrode} calculated using the PET model and the leading order electrolyte equations, (\ref{lyte1})-(\ref{lyte4}), at 8C discharge rate. In the upper panels a single snapshot in time (at the end of discharge) is shown because the electrolyte approaches a steady-state rapidly and hence the profiles at earlier times look very similar to those shown. Concentration of Lithium on the electrode particle surfaces (bottom left) and the potential in the solid (bottom right) across the $\text{Li}(\text{Ni}_{0.4}\text{Co}_{0.6})\text{O}_2$ electrode calculated using the full Newman PET at 8C discharge rate. Arrows indicates the direction of increasing time {and plots are made at 20 evenly spaced times between 0 and 0.10 hrs}. The leading order electrolyte equations are used to evaluate the first order terms in the corrected SPM, see (\ref{vapprox_1}).}
\label{SPM_PET_LOP}
\end{figure}

Simulations of both the basic SPM and its more accurate extension, the corrected SPM, are compared to simulations of the full PET model in figure \ref{SPM_PET_plot} \{as discharge curves showing cell voltage $V(t)$ plotted against time $t$ in the left-hand panels and $V(t)$ plotted against capacity in the right-hand panel. We examine three different electrode chemistries, namely, (i) a graphite anode $\text{LiC}_6$, (ii) an NMC nickel-cobalt oxide cathode $\text{Li}(\text{Ni}_{0.4}\text{Co}_{0.6})\text{O}_2$ and (iii) an LFP lithium-iron phosphate $\text{LiFePO}_4$ cathode. The parameter values for the former two of these are taken from the work of Ecker et. al \cite{Ecker2015a,Ecker2015b} whilst the latter are taken from Ranom \cite{ranom15}. {The open circuit voltages $\ueq(c_s)$  for all three materials are fitted to data from \cite{Ecker2015a,Ecker2015b,ranom15} in figure \ref{Ueq}, while the electrolyte diffusion coefficient $D(c)$ and the electrolyte conductivity $\kappa(c)$, for all three electrodes, is fitted to data in \cite{Ecker2015a,Ecker2015b} in figure \ref{Dliq} (top). Finally the active material diffusion coefficient $D_s(c_s)$ for the graphite and NMC particles are fitted to data from \cite{Ecker2015a,Ecker2015b} in figure \ref{Dliq} (bottom). Diffusion in the LFP nanoparticles is modelled by a linear diffusion equation with diffusion coefficient $D_s=9 \times 10^{-14} $m$^2$s$^{-1}$ as in Table~\ref{tbl:params} (this is so fast and the particles so small that diffusion is effectively instantaneous). All other parameter values are tabulated in Tables~\ref{tbl:params} and \ref{tbl:params_sep}.} {Figure \ref{SPM_PET_LOP} shows electrolyte variables ($c$ upper left panel and $\phi$ upper right panel) plotted as a function of distance $x$ across the half cell in addition to lithium concentration on the particle surfaces $c_s|_{r=R}$ (lower left panel) and electrode potential $\phi_s$ (lower right panel), all for an 8C discharge of the $\text{Li}(\text{Ni}_{0.4}\text{Co}_{0.6})\text{O}_2$ half-cell.}

{We observe that the basic SPM, reliably reproduces the results of the full PET across a range of chemistries provided that the rates are relatively low, i.e., less than 1C or so. As discharge rates increase beyond around 1C the accuracy of the SPM deteriorates. However, if the correction terms are accounted for, the range of accuracy of the SPM can be expanded significantly. In particular, for graphite, the results of full PET and the corrected SPM are almost indistinguishable even up to the relatively aggresive rate of 12C. For NMC and LFP electrodes, excellent agreement between the PET and corrected SPM is maintained until around 16C and 4C respectively. Beyond these values, even the corrected SPM begins to be noticeably different from the PET model, but, still performs far better than the usual SPM approximation.}

{As we will discuss in more detail in \S\ref{asy} the reliability of the corrected SPM is intimately linked to the characteristic size in the change of the overpotential of the electrode material as it is (de)lithiated. In particular it is surprising that good agreement is obtained between the corrected SPM and the PET model in the case of the LFP electrode because formally the SPM derivation, as we shall show in \S \ref{asy}, should not apply to a material with such a flat discharge curve.}

\subsection{The Double Particle Model and comparison to the Newman model \label{DPM}}
{If instead of an electrode comprised of just one size of electrode particle we consider an electrode composed of two sizes of particle distributed in space so that 
\be
R(x)=\left\{ \begin{array}{lll} R^{(a)} & \mbox{for} & 0\leq x < \alpha \\
R^{(b)}  & \mbox{for} & \alpha \leq x < L \end{array} \right. ,
\ee
we find that the leading order equations \eqref{appr1}-\eqref{appr3} reduce to the double particle model
\be
\left. \begin{array}{c} \ds \frac{\partial c^{(a)} _{s}}{\partial t}=\frac{1}{r^2}\frac{\partial}{\partial r}\left(r^2 D_s(c^{(a)} _{s})\frac{\partial c^{(a)} _{s}}{\partial r}\right)   \quad \mbox{in} \quad 0 \leq r \leq R^{(a)} \\*[4mm]
c^{(a)}_{s} \,\, \mbox{bounded on} \,\, r=0, \quad \left.  c_{s}^{(a)} \right|_{r=R^{(a)} } =\C(t)
\end{array} \right\}, \qquad c^{(a)}_{s}|_{t=0}=c_{s,init}, \label{dpm1} \\
\left. \begin{array}{c} \ds  \frac{\partial c^{(b)} _{s}}{\partial t}=\frac{1}{r^2}\frac{\partial}{\partial r}\left(r^2 D_s(c^{(b)} _{s})\frac{\partial c^{(b)} _{s}}{\partial r}\right)  \quad \mbox{in} \quad 0 \leq r \leq R^{(b)},\\
c^{(b)}_{s} \,\, \mbox{bounded on} \,\, r=0,  \quad \left.  c_{s}^{(b)} \right|_{r=R^{(b)} } =\C(t),
\end{array} \right\}, \qquad c^{(b)}_{s}|_{t=0}=c_{s,init},\label{dpm2} \\
\int_0^{\alpha} b(x)\left. \left( D_s(c^{(a)} _{s}) \frac{\partial c^{(a)} _{s}}{\partial r} \right) \right|_{r=R^{(a)} }  dx+\int_{\alpha}^L b(x)\left. \left( D_s(c^{(b)} _{s}) \frac{\partial c^{(b)} _{s}}{\partial r} \right) \right|_{r=R^{(b)} }  dx =  \frac{I(t)}{AF}.\label{dpm3}
\ee
Once these have been solved, the functions $G(t)$ and $\C(t)$ can, once again, be used as inputs in the electrolyte model \eqref{lyte1}-\eqref{lyte4} which we can then solve to obtain the data necessary to compute the voltage according to the corrected SPM, \ie via \eqref{vapprox_1}.}

\begin{figure} \centering
\includegraphics[width=0.495\textwidth]{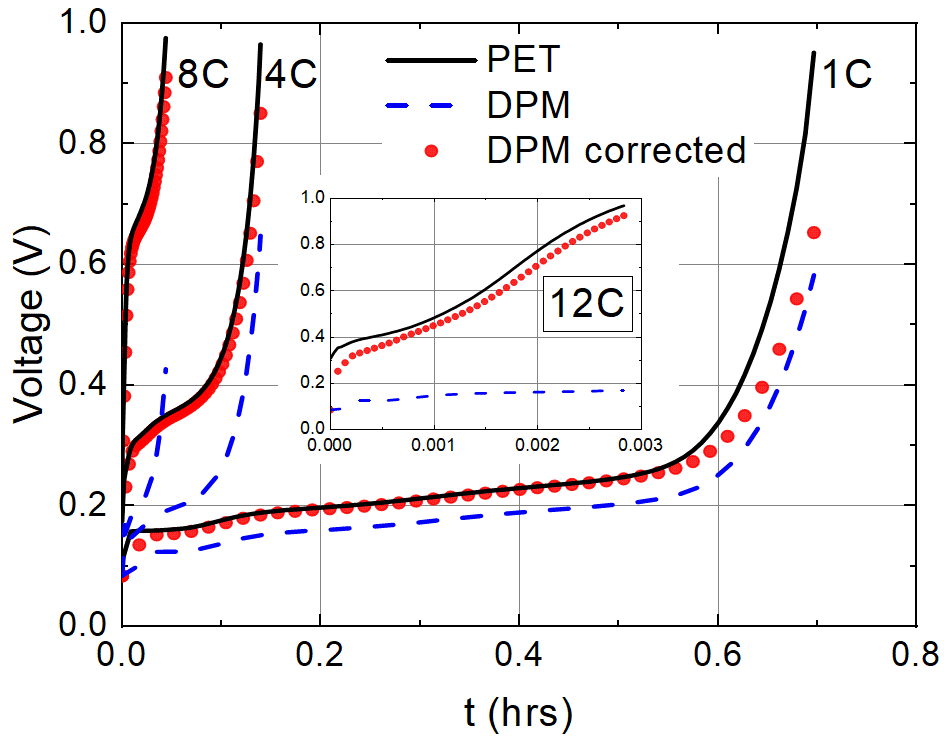}
\includegraphics[width=0.495\textwidth]{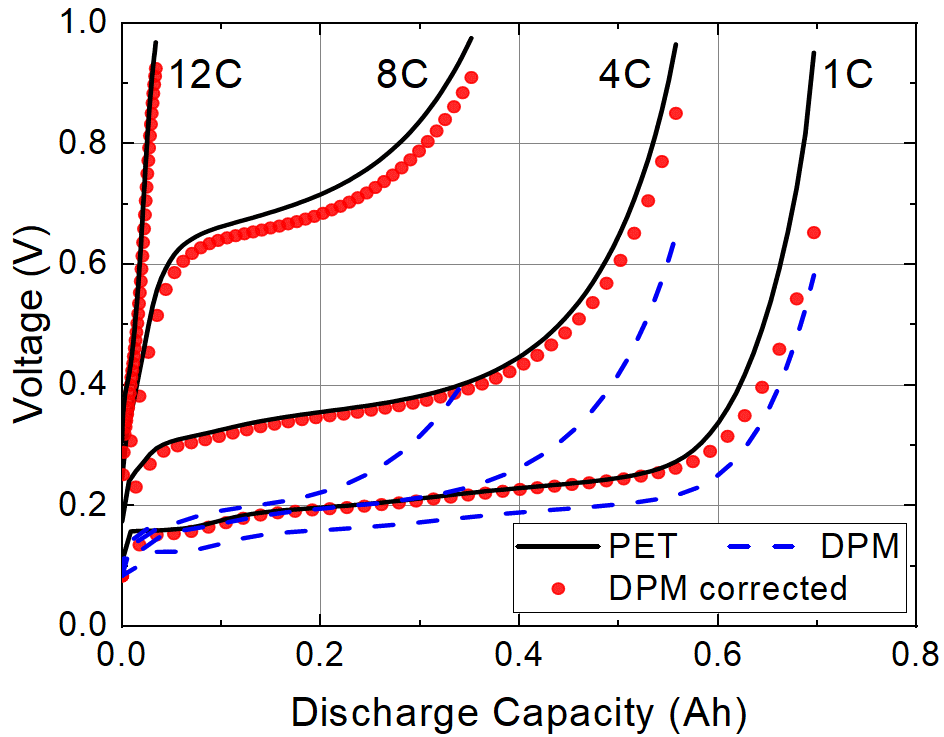}\\
\medskip
\includegraphics[width=0.495\textwidth]{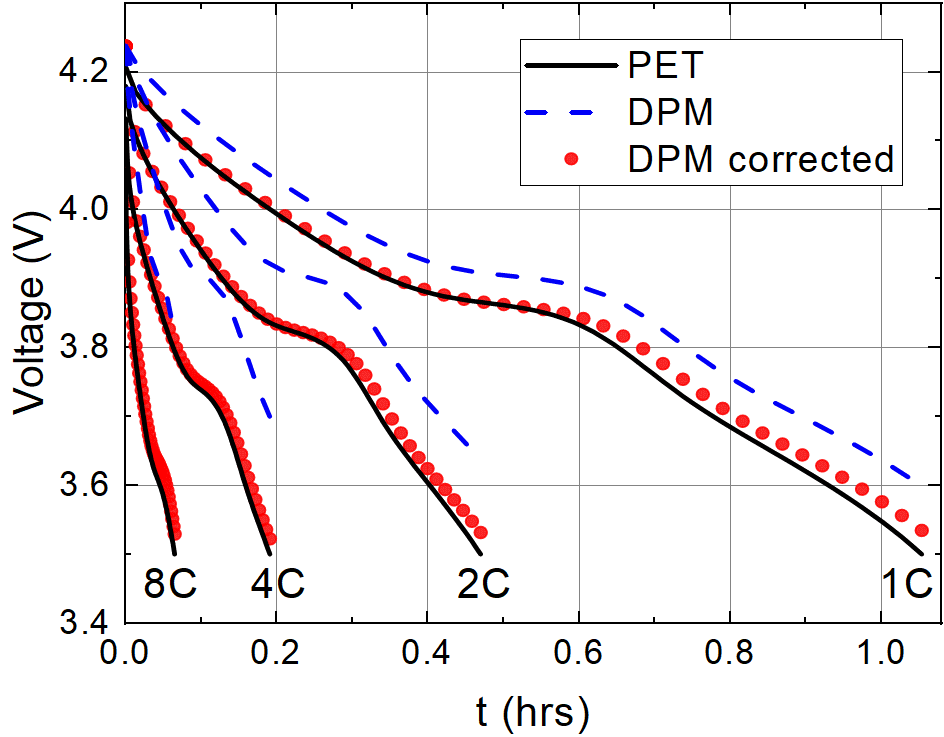}
\includegraphics[width=0.495\textwidth]{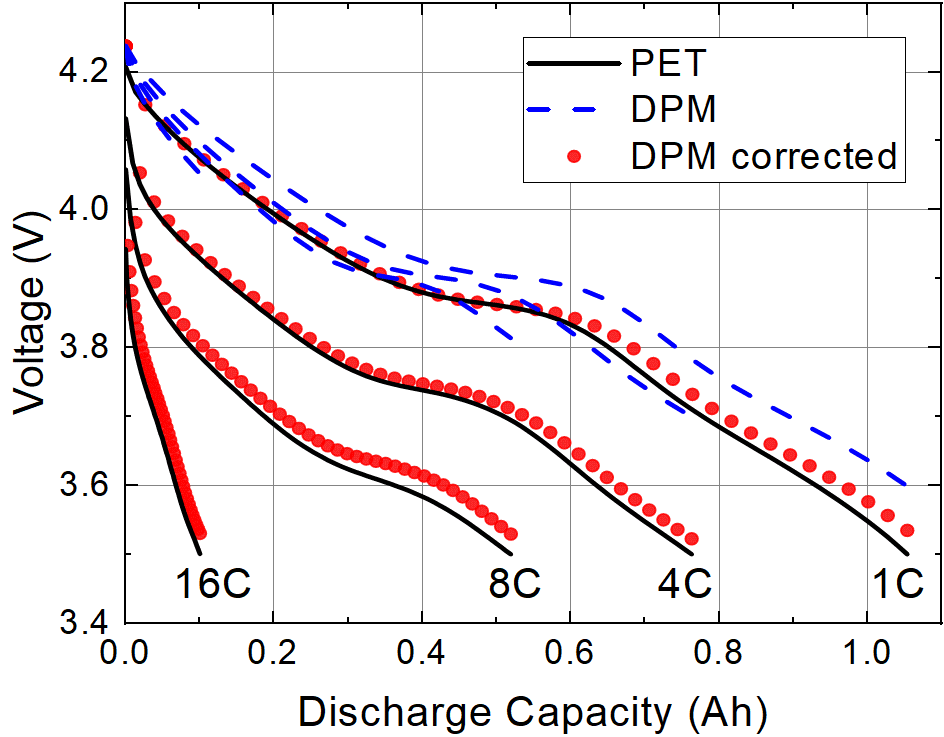}\\
\medskip
\includegraphics[width=0.495\textwidth]{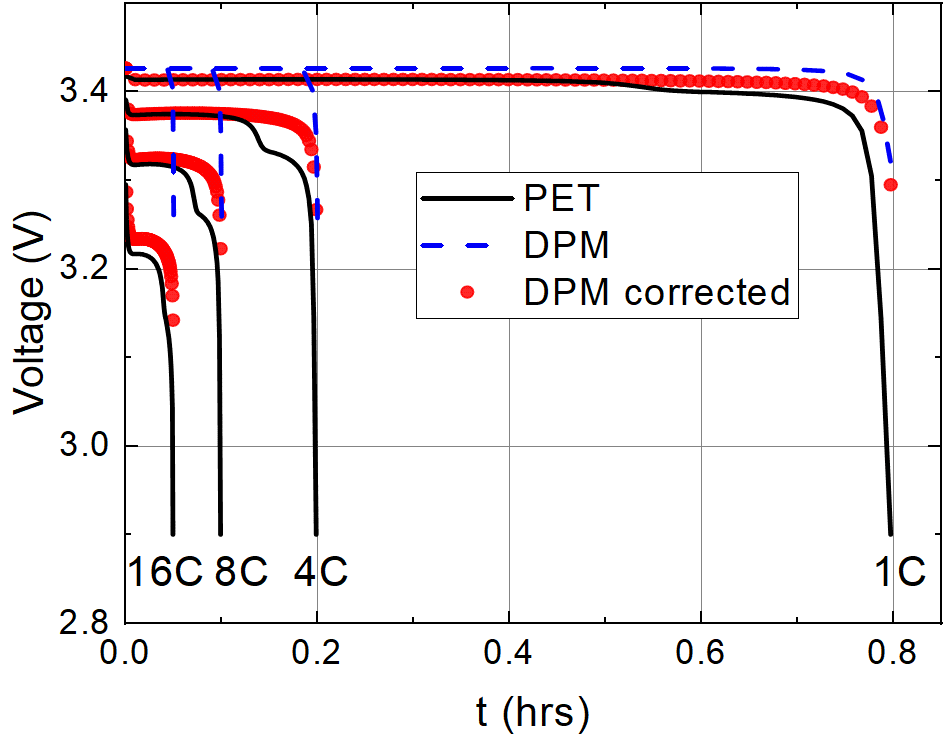}
\includegraphics[width=0.495\textwidth]{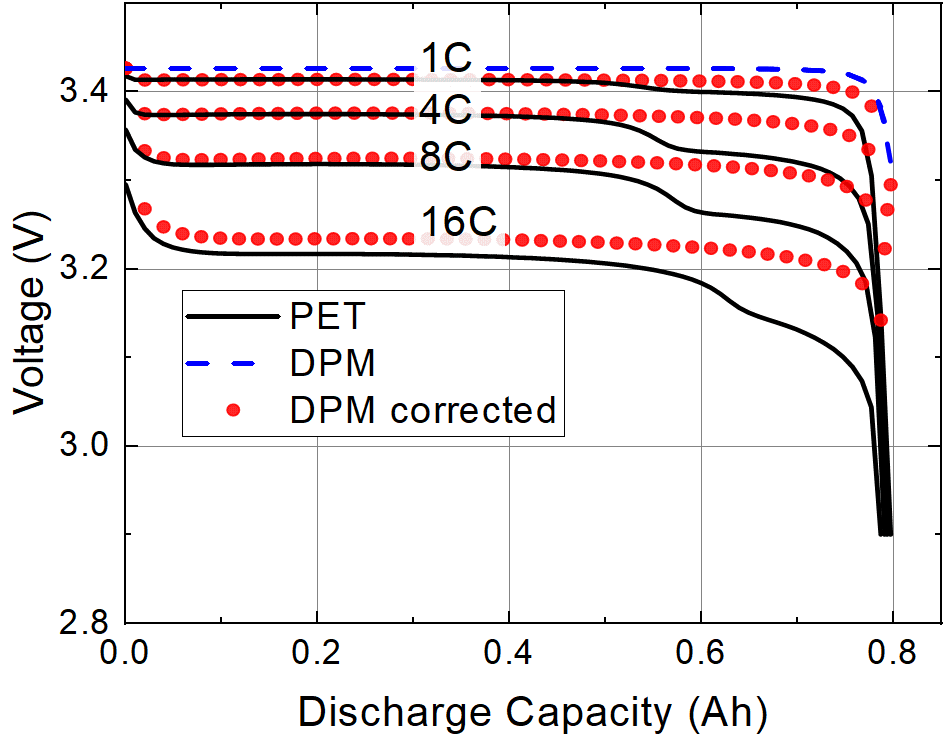}\\
\caption{Cell potentials $V$ calculated using the full PET model, DPM and the corrected DPM for graphite ($\text{LiC}_6$) anode (top row), $\text{Li}(\text{Ni}_{0.4}\text{Co}_{0.6})\text{O}_2$ cathode (middle row) and LFP ($\text{LiFePO}_4$) electrode (bottom row) at different discharge rates. {Results are shown against both time (left) and discharge capacity (right).}}
\label{DPM_PET_plot}
\end{figure}

\begin{figure} \centering
\includegraphics[width=0.495\textwidth]{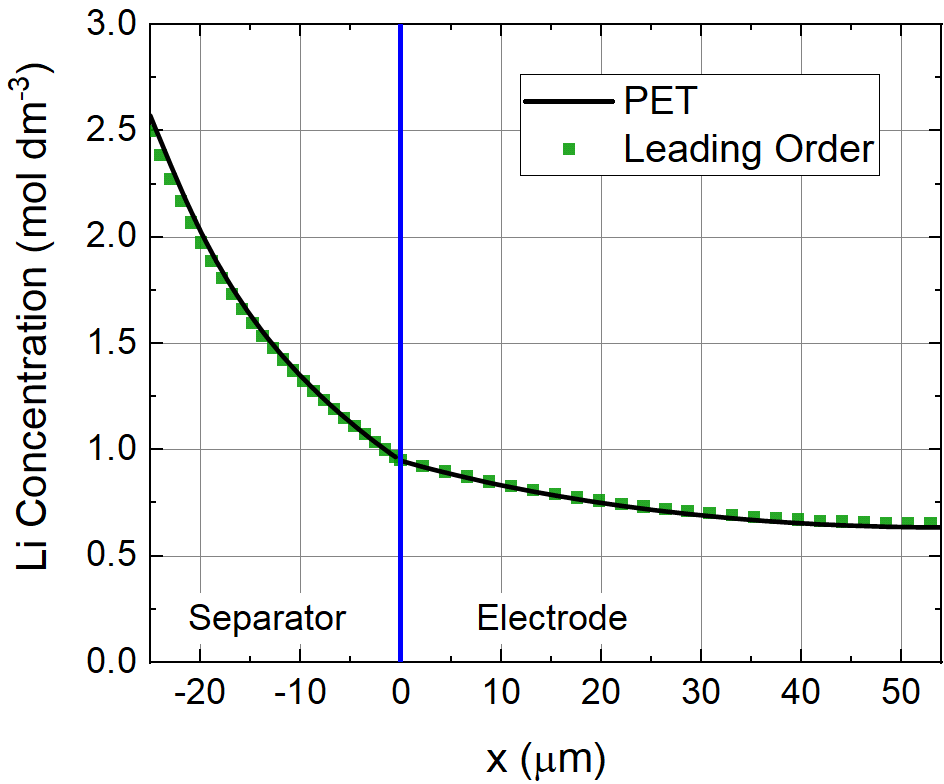}
\includegraphics[width=0.495\textwidth]{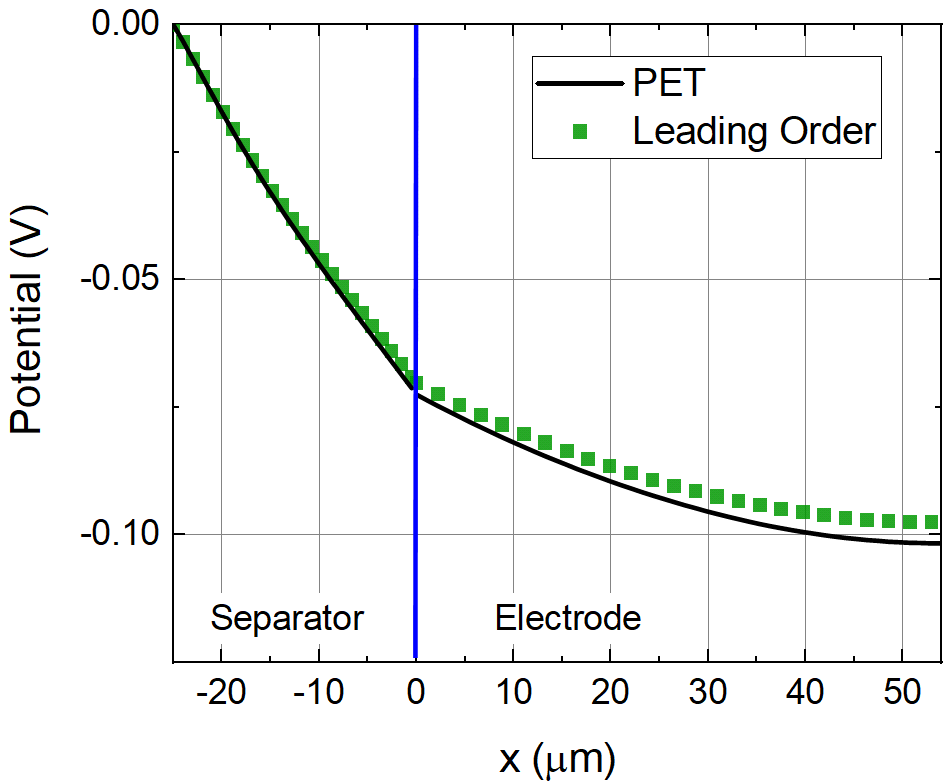}\\
\medskip
\includegraphics[width=0.495\textwidth]{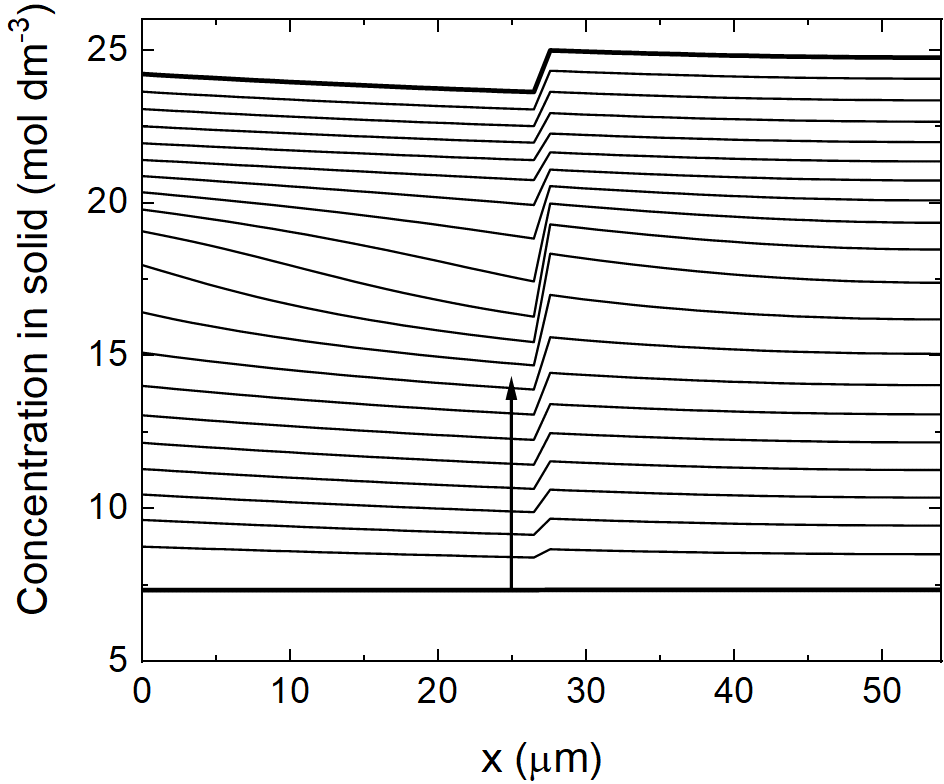}
\includegraphics[width=0.495\textwidth]{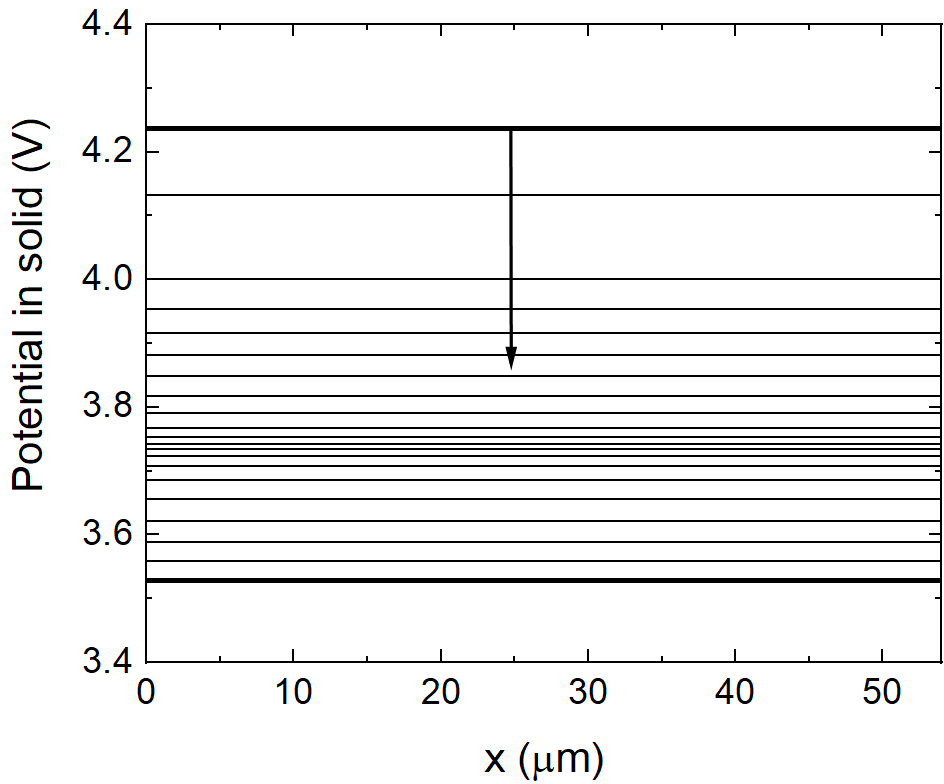}\\
\caption{Concentration of Li$^{+}$ ions (top left) and potential (top right) in the electrolyte {across a half-cell with an $\text{Li}(\text{Ni}_{0.4}\text{Co}_{0.6})\text{O}_2$ electrode formed from particles of two different sizes and} calculated using the full PET model and the leading order approximation at 4C discharge rate. In the upper panels a single snapshot in time (at the end of discharge) is shown because the electrolyte approaches a steady-state rapidly and hence the profiles at earlier times look very similar to those shown. Concentration of Li on the particle surfaces (bottom left) and potential in solid (bottom right) across an $\text{Li}(\text{Ni}_{0.4}\text{Co}_{0.6})\text{O}_2$ electrode with two different particle sizes calculated using the full PET model at 4C discharge rate. Arrows indicates the direction of increasing time {and plots are made at 20 evenly spaced times between 0 and 0.19 hrs}.}
\label{DPM_PET_LOP}
\end{figure}

{Figures \ref{DPM_PET_plot} compare the results of this double particle model, both without the correction term (DPM) and with the correction term (DPM corrected), to the PET model. Here we take $\alpha=L/2$ and $R^{(a)}=4 R^{(b)}$, $b^{(a)}=b^{(b)}/4$ in all three cases while $R^{(b)}$ and $b^{(b)}$ are given by the values of $R_0$ and $b_0$ in Table \ref{tbl:params} for the appropriate chemistries. Figure \ref{DPM_PET_LOP} shows the equivalent electrolyte variables ($c$ upper left panel and $\phi$ upper right panel) plotted as a function of distance $x$ across the half cell in addition to lithium concentration on the particle surfaces $c_s|_{r=R(x)}$ (lower left panel) and electrode potential $\phi_s$ (lower right panel), all for an 8C discharge of the graded $\text{Li}(\text{Ni}_{0.4}\text{Co}_{0.6})\text{O}_2$ half-cell.}

{We note that the inclusion of the first order correction terms into the model (\ie upgrading to the corrected DPM) in figure \ref{DPM_PET_plot}  significantly improves the agreement with the full PET simulation over that with the simple DPM throughout the range of (dis)charge rates and chemistries we explored . Notably in the case of the LFP half-cell the agreement of DPM corrected with the PET simulations is not as good as it was for the electrode with single particle size. However some discrepancy is to be expected between the corrected DPM model and the PET model for an electrode formed from an active material with such a flat discharge curve.}


\section{\label{sec:nondim}Nondimensionalisation of the model} Before applying asymptotic methods to derive the approximate models described in \eqref{appr1}-\eqref{lyte4} from the underlying Newman model \eqref{dim1}-\eqref{zap} we must non-dimensionalise the PET model.  A key quantity, that will be used later for our temporal scaling, is $\tau$ the characteristic timescale for cell (dis)charge, i.e., the characteristic timescale over which an electrode can sustain a current of size, $\hat{I}$. This is 
\be \label{deftau}
\tau = \frac{AL F \csm \hat{b} \hat{R}}{\hat{I}},
\ee
where $\hat{I}$, $\hat{b}$ and $\hat{R}$ are typical sizes of the $I$, $b$ and $R$ respectively. The timescale $\tau$ can be directly assessed by examining typical C-rates used for battery discharge. Here, we will examine a range of C-rates from the relatively mild (1C, corresponding to a half cycle time of 1hour) to the relatively aggresive (16C, corresponding to a half cycle time of {3.75} minutes).

The spatial coordinate $x$ is scaled based on the width of the electrode $L$. The quantities $c$, $c_s$, $R$, $D$, $D_{s}$, $\kappa$, $I$, ${\mathcal B}$, $\sigma$ and $b$ are scaled with their typical values, namely $c_{\text{init}}$, $\csm$, $\hat{R}$, $\hat{D}$, $\hat{D}_{s}$, $\hat{\kappa}$, $\hat{I}$, $\hat{\mathcal B}$, $\hat{\sigma}$ and $\hat{b}$ respectively. The BV reaction rate will be scaled based on the average flux required through particle surfaces to sustain a current of size $\hat{I}$. The scaling for the effective ionic flux is based on the size of the diffusive flux carried by a concentration of size ${c}_{\text{init}}$ over lengths of size $L$ with an effective diffusivity of size $\hat{\mathcal B} \hat{D}$. 

It remains to specify scalings for the various different potentials. One natural scale is that of the variation of the electric potential within the electrolyte. The conductivities of typical electrolytes are chosen so that only small variations in potential, on the order of the thermal voltage $RT/F=26$mV, are sufficient to carry the requisite current densities suggesting that an appropriate scaling for $\phi$ is the thermal voltage. A second natural potential scale, which we will henceforth refer to as the characteristic half cell voltage, $\mathcal U$, is the size of the difference between the overpotentials of a fully lithiated electrode particle and that of a fully delithiated one. The overpotential of graphite varies between around 0V at full lithiation and 1V at full delithiation whilst the various metal oxides used in cathodes, \eg NMC, exhibit variations in their overpotentials of 3V at full lithation and 4V at full delithitaion. We proceed on the basis that a typical value the characteristic cell potential, $\mathcal U$ can be taken to be on the order of 1V. A notable exception if LFP which exhibits a extremly flat discharge curve for which $\mathcal U \leq 26$mV. In order that the reaction rates on the particle surfaces are of the requisite size to satitate a current demand of size $\hat{I}$ we should scale both the electrode overpotential, $\ueq$, and the solid electrode potential, $\phi_s$, with the characteristic cell voltage. It is the vast disparity between the thermal voltage and the characteristic cell voltage that gives rise to the large value of the dimensionless parameter $\lambda$ and facilitates the asymptotic analysis that justifies the SPM derivation. In summary, our scalings for the problem are
\begin{align} \label{Rescalings 1} 
&t=\tau t^* & &x=Lx^* & &{r=\hat{R} r^*} & &D=\hat{D} D^* \\
&D_{s}=\hat{D}_{s} D_{s}^*  & & \kappa=\hat{\kappa} \kappa^* & &c=c_{\text{init}} c^* & &{c_{s}=\csm c_{s}^*} \\&V={\mathcal U} V^* & &\phi=\frac{RT}{F} \phi^* & &\phi_s={\mathcal U} \phi_s^* & &j=\frac{\hat{I}}{A} j^* \\
&j_s=\frac{\hat{I}}{A} j_s^* & &I=\hat{I} I^* & &G=\frac{\hat{I}}{ALF\hat{b}} G^* & &\eta=\frac{RT}{F} \eta^* \\ 
&\ueq={\mathcal U} \ueq^* & &\Fms=\frac{\hat{\mathcal B} \hat{D} c_{\text{init}}}{L} \Fm^* & &\mathcal{B}=\hat{\mathcal B} \mathcal{B}^* & & b = \hat{b} b^* \\
& I = \hat{I} I^* && R = \hat{R} R^* && \sigma = \hat{\sigma} \sigma^*. \label{Rescalings 2}
\end{align}
The non-dimensionalisation gives rise to the following dimensionless quantities that characterise the system: 
\begin{align}
\label{capital_dless}
&\mathcal{N}=\frac{L^{2}}{\tau \hat{\mathcal{B}} \hat{D}}, & &\Gamma=\frac{\hat{I} L}{Ac_{\text{init}} \hat{D} \hat{\mathcal{B}} F}, & &\lambda=\frac{{\mathcal U} F}{RT}, & &\mathcal{Q}=\frac{\hat{R}^{2}}{\tau \hat{D}_{s}},\\ 
&\Upsilon =\frac{k {c_{s,\text{init}}}^{1/2} \csm A L F \hat{b}}{\hat{I}}, & &\Theta=\frac{\hat{\sigma} RTA}{L\hat{I}F},  & &\mathcal{P}=\frac{\hat{\mathcal{B}} \hat{\kappa} RTA}{L \hat{I} F}, && \LL_s = \frac{L_s}{L}\\
& \csin = \frac{c_{s,\text{init}}}{\csm}.\label{capital_dless2} 
\end{align}
Of those parameters whose meaning is not self-evident from their definition; $\mathcal{N}$ is the timescale for diffusion in the electrolyte over the timescale for cell discharge, $\Gamma$ is the drift flux over the diffusive flux, $\mathcal Q$ is the timescale for diffusive transport inside the electrode particles over the timescale for cell discharge, $\Upsilon$ is the total amount of lithium intercalated into the active material per second over the current, $\Theta$ is the electronic conductivity of the solid scaffold over the characteristic conductivity of the electrode material, $\mathcal{P}$ is the ionic conductivity over the characteristic conductivity of the electrode material.

\bgroup
\def\arraystretch{1.25}  
\begin{table}
\begin{center}
\caption{Parameter values used in the model for different chemistries}
\label{tbl:params}
\begin{tabular}{|c|c|ccc|} 
\hline
Parameter & 
Units & 
\makecell{Graphite ($\text{LiC}_6$) \\ \cite{Ecker2015b,Ecker2015a}} & 
\makecell{$\text{Li}(\text{Ni}_{0.4}\text{Co}_{0.6})\text{O}_2$ \\ \cite{Ecker2015b,Ecker2015a}} & 
\makecell{LFP ($\text{LiFePO}_4$) \\ \cite{ranom15,kang11}} \\
\hline
Electrode thickness, $L$ & $\mu$m & 74 & 54 & 62 \\
Electrode particle radius, $R_0$ & $\mu$m & 13.7 & 6.5 & 0.05 \\
Electrode cross-section area, $A$ & m$^2$ & $8.585 \times 10^{-3}$ & $8.585 \times 10^{-3}$ & $10^{-4}$ \\
Vol. fraction electrolyte, $\epsilon_l$ & -- & 0.329 & 0.296 & 0.4764 \\
\makecell{Brunauer-Emmett-Teller \\ surface area, $b_0 = 3(1-\epsilon_l)/R_0$} & $m^{-1}$ & $1.469 \times 10^5$ & $3.249 \times 10^5$ & $3.142 \times 10^7$ \\
Conductivity in solid, $\sigma_0$ & S\,m$^{-1}$, & 14.0 & 68.1 & 0.5 \\
\makecell{Permeability factor of \\ electrolyte, $\mathcal{B}_0$} & -- & 0.162 & 0.153 & 0.329 \\
Reaction rate constant, $k$ & m$^{2.5}$s$^{-1}$mol$^{-0.5}$ & $2.333 \times 10^{-10}$ & $5.904 \times 10^{-11}$ & $3 \times 10^{-12}$ \\
\makecell{Maximum concentration of \\ Li ions in solid, $\csm$} & mol\,m$^{-3}$ & 17715.6 & 28176.4 & 18805 \\
Transference number, $t^+$ & -- & 0.26 & 0.26 & 0.3 \\
1C current draw, $I_0$ & A & -0.15625 & 0.15625 & 0.0015 \\
Contact resistance, $R_c$ & $\Omega$ & 0 & 0 & 0 \\
Absolute temperature, $T$ & $K$ & 298.15 & 298.15 & 298.15 \\
\makecell{Typical concentration of Li \\ in liquid, $c_{init}$} & mol\,m$^{-3}$ & 1000 & 1000 & 1000 \\
Typical diffusivity liquid, $D_0$ & m$^2$s$^{-1}$ & $2.594 \times 10^{-10}$ & $2.594 \times 10^{-10}$ & $2.594 \times 10^{-10}$ \\
Typical conductivity liquid, $\kappa_0$ & S\,m$^{-1}$ & 1.0 & 1.0 & 1.0 \\
Diffusivity in solid, $D_s$ & m$^2$s$^{-1}$ & Fig.\ref{Dliq} & Fig.\ref{Dliq} & $9 \times 10^{-14}$ \\
Typical diffusivity in solid, $\mathcal{D}_0$ & m$^2$s$^{-1}$ & $3 \times 10^{-14}$ & $10^{-13}$ & $9 \times 10^{-14}$ \\
Equilibrium potential, $\ueq$ & V & Fig.\ref{Ueq} & Fig.\ref{Ueq} & Fig.\ref{Ueq} \\
Characteristic cell voltage, $\mathcal{U}$ & V & {1.0} & {1.0} & {1.0} \\
{Discharge time scale (\ref{deftau})}, $\tau$ & s & $1.399 \times 10^{4}$ & $1.703 \times 10^{4}$ & $1.178 \times 10^{4}$ \\
\hline
\multicolumn{5}{|c|}{Derived dimensionless quantities (\ref{capital_dless})-(\ref{capital_dless2})} \\
\hline
$\mathcal{N}={L^{2}}/{(\tau \hat{\mathcal{B}}\hat{D})}$ & -- & 0.0093 & 0.0043 & 0.0038 \\
$\Gamma={I_0 L}/{(Ac_{\text{init}}\hat{D}\hat{\mathcal{B}}F)}$ & -- & 0.332 & 0.257 & 0.113 \\
$\Upsilon ={k c_{\text{init}}^{1/2} {\csm} A L F \hat{b}}/{\hat{I}}$ & -- & 7.53 & 4.89 & 22.4 \\
$\Theta={\hat{\sigma}RTA}/{(L \hat{I}F)}$ & -- & 267 & 1780 & 13.8 \\
$\mathcal{P}={\hat{\mathcal{B}}\hat{\kappa}_RTA}/{(L \hat{I}F)}$ & -- & 3.1 & 4.0 & 9.1 \\
{$\mathcal{Q}={\hat{R}}^2/{\tau \hat{D}_s}$} & -- & 0.447 & 0.0248 & $2.36 \times 10^{-6}$ \\
\hline
\end{tabular}
\end{center}
\end{table}
\egroup

\bgroup
\def\arraystretch{1.25}  
\begin{table}
\begin{center}
\caption{Parameters of the separator used in the model}
\label{tbl:params_sep}
\begin{tabular}{|c|c|c|} 
\hline
Parameter & Units & Separator \cite{Sri04}\\
\hline
Thickness, $L^{sep}$ & $\mu$m & 25 \\
Volume fraction of electrolyte, $\epsilon_l^{sep}$ & -- & 0.55 \\
Permeability factor of electrolyte, $\mathcal{B}_0^{sep}$ & -- & 0.408 \\
\hline
\end{tabular}
\end{center}
\end{table}
\egroup

{\subsection{\label{nondim}The Half-Cell Dimensionless Model} On dropping the stars from the dimensionless variables the dimensionless problem reads as follows. In the region $-\LL_s<x<0$ the (dimensionless) electrolyte equations are
\be 
\epsilon_l(x) \mathcal{N}\frac{\partial c}{\partial t}+\frac{\partial \Fm}{\partial x}=0, \quad \Fm=-{\mathcal B}(x) D(c)\frac{\partial c}{\partial x}-\Gamma(1-t^+)j, \label{nd_1}\\
\frac{\partial j}{\partial x}=\left\{ \begin{array}{llc}0 & \mbox{for} & -\LL_s<x<0\\
b(x) G(x,t)& \mbox{for} & 0<x<1 \end{array} \right. , \label{nd_2}\\
j=-\mathcal{P}\kappa(c) {\mathcal B}(x) \left( \frac{\partial \phi}{\partial x}-2\frac{1-t^+}{c}\frac{\partial c}{\partial x}\right),  \label{nd_3}
\ee
and satisfy the boundary conditions
\be
\begin{array}{llll}
j|_{x=-\LL_s} = I(t),  & \Fm|_{x=-\LL_s} = 0, & \phi|_{x=-\LL_s} =0,
 & {\Fm}|_{x=1} = 0. \label{nd_4}
 \end{array} 
 \ee
 These couple to the (dimensionless) electrode equations in the region $0<x<1$ 
 \be
 \label{nd_5}
j_s=-\lambda \Theta\sigma \frac{\partial \phi_s}{\partial x}, \qquad \frac{\partial j_s}{\partial x}=-b(x) G(x,t), \\
\label{nd_6} 
G(x,t)=2 \Upsilon c^{1/2} \left( c_{s}\rvert _{r=R(x)} \right)^{1/2}\left( 1-c_s\rvert _{r=R(x)} \right)^{1/2}  \sinh \left(\frac{ \eta}{2}\right), \\ 
 \eta=\lambda \left( \phi_s- \ueq(c_s|_{r=R(x)}) \right) -\phi,  \label{nd_7}\\
\label{nd_8} \Q \frac{\partial c_s}{\partial t}=\frac{1}{r^2}\frac{\partial}{\partial r}\left(r^2 D_s(c_s)\frac{\partial c_s}{\partial r}\right),\quad \mbox{for} \ \ 0\leq r \leq R(x),\\
\label{nd_9} c_s \,\, \mbox{bounded on} \,\, r=0,  \quad -D_s \frac{\partial c_s}{\partial r}\bigg\rvert_{r=R(x)}= \Q G.
\ee
which in turn satisfy two boundary conditions in $x$, namely
\be
\label{hc_bc2}  j_s|_{x=0} = 0, \qquad  j_s|_{x=1} = I(t), \qquad V(t)=\phi_s|_{x=1}.
\ee
The final condition in (\ref{hc_bc2}) serves to determine the half-cell voltage $V(t)$ from the solution to the problem. Initial conditions corresponding to a half-cell which is initially at equilibrium are 
\be
\begin{array}{cccc}
\label{nd_ic}
c\rvert_{t=0}=1, & c_s\rvert_{t=0}= \csin.
\end{array}
\ee
where $\csin$ is uniform throughout the cell (\ie is independent of both $x$ and $r$). }

\paragraph{Features of the model}
Some helpful features of the model can be derived by taking the sum of equations (\ref{nd_2}) and (\ref{nd_5}b) in the region $0<x<1$, integrating the results and applying the boundary conditions (\ref{nd_4}d) and (\ref{hc_bc2}b). We arrive at
\be \label{curr_cons}
j+j_s = I(t).
\ee
Thus, the total (both ionic and electronic) current density is uniform throughout the electrode. Furthermore, on integrating (\ref{nd_2}) through the thickness of the electrode ($-\LL_s<x<1$) and applying the boundary conditions (\ref{nd_4}a) and (\ref{nd_4}d)  we obtain the integral condition
\be \label{totinterc}
\int_{0}^{1} b(x) G(x) dx = -I(t).
\ee
This amounts to the observation that the total amount of charge (de)intercalating within the electrode is in balance with the charge being deliver to/supplied from the external circuit.

\section{Asymptotic reduction: derivation of single and multiple particle models \label{asy}}
{In this section we systematically derive both the basic SPM and the corrected SPM for the half cell from the dimensionless PET model \eqref{nd_1}-\eqref{hc_bc2} using asymptotic methods in the large $\lambda$ limit. In addition to deriving these SPMs we also derive natural extensions that describe electrodes with more than one particle size and/or chemistry. {This analysis can easily be extended to a full-cell, with two porous electrodes, but we do not do this here.} Formally we investigate the distinguished limit in which $\lambda \ra \infty$ and all other dimensionless parameters (\ie $\N$, $\Gamma$, $\LL_s$, $\PP$, $\Theta$, $\Upsilon$, $\csin$ and $\Q$) are order 1. }

{The key observation that leads to the SPM is that in order that the dimensionless current $I(t)$ is of size $O(1)$ the total amount of intercalation throughout the electrode must also be size $O(1)$ (see (\ref{totinterc})). In turn this requires that, the dimensionless overpotential, $\eta$ is $O(1)$ throughout the electrode which leads to the condition that 
\be \label{SPMkey}
\phi - \lambda \left( \phi_s -  \ueq(c_s|_{r=R(x)}) \right) = O(1).
\ee
Since $I(t)$ is $O(1)$ it follows that  both the dimensionless current densities $j$ and $j_s$, within the electrolyte and electrode respectively, are $O(1)$. The condition that $j=O(1)$ and equation \eqref{nd_4} imply that $\phi=O(1)$ while the condition that $j_s=O(1)$ means that gradients in $\phi_s$  are $O(1/\lambda)$. This leads us to the following asymptotic expansion:
\be
\begin{array}{ccccc}
\ds \phi_s =  \phi_{s,0}+ \frac{1}{\lambda}\phi_{s,1} + \cdots, & \ds V=V_0(t)+\frac{1}{\lambda} V_1(t) + \cdots, & \ds c_{s}=c_{s,0}+\frac{1}{\lambda}c_{s,1} \cdots, \\*[8mm]
\ds j_s=j_{s,0}+\frac{1}{\lambda}j_{s,1} \cdots, &\ds \eta_i = \eta_{0} +\frac{1}{\lambda}\eta_{1} \cdots, &\ds G_i = G_{0} + \frac{1}{\lambda}G_{1} \cdots, \\*[4mm]
\phi = \phi_0 + \cdots, &  \ds c=c_0+\cdots & \ds j=j_0+\cdots, \\
& \Fm=\Fmz+ \cdots.
\end{array} \label{exp1}
\ee}

{\paragraph{The leading order problem.} The derivation of the leading order term presented here broadly follows that in the thesis of Ranom \cite{ranom15}.  On inserting the expansions (\ref{exp1}) into the equations (\ref{nd_5})-(\ref{hc_bc2}) and taking the leading order terms we obtain the following problem in $0 \leq x \leq 1$:
\be
\frac{\dd \phi_{s,0}}{\dd x}&=&0, \label{lo1} \\
\frac{\dd j_{s,0}}{\dd x}&=& -b(x) G_0(x,t), \quad j_{s,0}|_{x=0}=0, \quad  j_{s,0}|_{x=1}=I(t), \label{lo2} \\
\phi_{s,0}&=&\ueq\left( c_{s,0} |_{r=R(x)} \right), \quad V_0(t)= \phi_{s,0}|_{x=1}, \label{lo3} \\
\mathcal{Q} \frac{\partial c_{s,0}}{\partial t}&=&\frac{1}{r^2}\frac{\partial}{\partial r}\left(r^2 D_s(c_{s,0})\frac{\partial c_{s,0}}{\partial r}\right) \quad \mbox{in} \quad 0 \leq r \leq R(x),\label{lo4} \\
c_{s,0} &\,\, & \mbox{bounded on} \,\, r=0,  \quad D_s(c_{s,0}) \frac{\partial c_{s,0}}{\partial r}\bigg\rvert_{r=R(x)}= -\Q G_{0}(x,t), \label{lo5} \\
c_{s,0}|_{t=0} &=&\csin. \label{lo6} 
\ee
This system is solved by noting that the integral of \eqref{lo1} implies that $ \phi_{s,0}= \phi_{s,0}(t)$ and in turn that (\ref{lo3}a) implies that
\be
c_{s,0} |_{r=R(x)} = \ueq^{-1}\left(\phi_{s,0}(t) \right).  \label{lo7}
\ee
Integrating (\ref{lo2}a) between $x=0$ and $x=1$ and applying the boundary conditions (\ref{lo2}b)-(\ref{lo2}c) leads to the condition
\be
\int_0^1 b(x) G_0(x,t) dx= -I(t),  \label{lo8}
\ee
which is equivalent to the leading order term of the condition \eqref{totinterc}.
The leading order problem thus comprises the sequence of electrode particle problems \eqref{lo4}-\eqref{lo6} coupled to the conditions \eqref{lo7}-\eqref{lo8} with the leading order half-cell potential $V_0(t)$ being found from (\ref{lo3}b). In summary the leading order problem can be written in the form
\be
\Q\frac{\partial c_{s,0}}{\partial t}=\frac{1}{r^2}\frac{\partial}{\partial r}\left(r^2 D_s(c_{s,0})\frac{\partial c_{s,0}}{\partial r}\right)  \quad \mbox{in} \quad 0 \leq r \leq R(x),\label{lo10} \\
c_{s,0} \,\, \mbox{bounded on} \,\, r=0,  \quad c_{s,0}|_{r=R(x)}=\C_0(t),\label{lo11} \\
\int_0^1 b(x)\left. \left( D_s(c_{s,0}) \frac{\partial c_{s,0}}{\partial r} \right) \right|_{r=R(x)}  dx= \Q I(t),\label{lo12} \\
V_0(t)= \phi_{s,0}(t) = \ueq( \C_0(t)).\label{lo13} 
\ee
Here $\C_0(t)$ is chosen at each time step so that the integral condition \eqref{lo12} is satisfied. The leading order reaction rate $G_0(x,t)$ is determined from the solution to this problem via the condition
\be
G_0(x,t)= -\frac{1}{\Q} D_s(c_{s,0}) \frac{\partial c_{s,0}}{\partial r}\bigg\rvert_{r=R(x)}.
\ee}

{\paragraph{Remark.} The solution of the sequence of diffusion problems  \eqref{lo10}-\eqref{lo13} does not represent a significant saving if the electrode particle radii $R(x)$ vary continuously in $x$. However where $R(x)$ is piecewise constant a very major saving can be achieved because, rather than solving a continuum of diffusion problems \eqref{lo10}-\eqref{lo11} in $x$ we need only solve a finite number of such problems.}

{\subsection{The single particle model (SPM)}
In the case of a uniform particle size throughout the half cell the leading order equations simplify significantly because 
all the particles are identical and so, on writing $R=1$ (recalling that we have nondimensionalised $r$ with typical particle radius $\hat{R}$), equations \eqref{lo10}-\eqref{lo13} simplify to
\be
\Q\frac{\partial c_{s,0}}{\partial t}=\frac{1}{r^2}\frac{\partial}{\partial r}\left(r^2 D_s(c_{s,0})\frac{\partial c_{s,0}}{\partial r}\right)   \quad \mbox{in} \quad 0 \leq r \leq 1,\label{spm1} \\
c_{s,0} \, \, \mbox{bounded on} \,\,r=0,  \quad \left. \left( D_s(c_{s,0}) \frac{\partial c_{s,0}}{\partial r} \right) \right|_{r=1}=\frac{\displaystyle \Q I(t)}{\displaystyle \int_0^1 b(x) dx},\label{spm2} \\
c_{s,0}|_{t=0}=\csin,\qquad \mbox{and} \qquad
V_0(t)= \ueq( c_{s,0}|_{r=1}).\label{spm3} 
\ee
}

\subsection{\label{foasy}Calculating the first order correction term}
{Here we seek to calculate the first order correction $V_1(t)$ to the voltage across the half-cell.  We start by deriving an expression for the first order electrode potential $\phi_{s,1}$. Substituting  expansion \eqref{exp1} into \eqref{nd_5} yields the problem
\be
 \frac{\partial \phi_{s,1}}{\partial x}= -\frac{j_{s,0}(x,t)}{\Theta\sigma(x)}, \quad  \phi_{s,1}|_{x=1}=V_1(t). \label{pied}
\ee
and on solving for $j_{s,0}$ from \eqref{lo1} and for $\phi_{s,1}$ from \eqref{pied} we obtain the required expression 
\be
j_{s,0}(x,t)= \int_0^x b(x')G_0(x',t) dx', \quad \phi_{s,1}(x,t)= V_1(t)+\int_x^1 \frac{j_{s,0}(x',t) }{\Theta\sigma(x')} dx'. \label{phis1}
\ee
Notably this formula for $\phi_{s,1}$ depends upon $V_1(t)$, the quantity that we seek.}

{\paragraph{Solvability condition.} We obtain a solvability condition, that may be used to determine $V_1(t)$, by writing down the first order expansion of the electrode current conservation equation (\ref{nd_5}b) and its boundary conditions (\ref{hc_bc2}a-b):
\be
\frac{\partial j_{s,1}}{\partial x}= -b(x) G_1(x,t), \quad j_{s,1}|_{x=0}=0, \quad \quad j_{s,1}|_{x=1}=0.
\ee
Integration of this equation, between $x=0$ and $x=1$, and application of  the boundary conditions leads to the solvability condition
\be
\int_0^1 b(x) G_1(x,t) dx=0, \label{solv_cond}
\ee
which is equivalent to the first order terms in \eqref{totinterc}. It remains to determine an appropriate expression for $G_1$ that can be substituted into the solvability condition (\ref{solv_cond}). This is accomplished by substituting the expansion \eqref{exp1} into the boundary condition (\ref{nd_9}b) and proceeding to first order; a procedure that yields the result
\be
G_1(x,t)=- \left. \frac{1}{\Q} \frac{\dd}{\dd r} (D_s(c_{s,0}) c_{s,1})  \right|_{r=R(x)} \label{G1expr}
\ee
In order to find determine the right-hand side of this expression we need to solve the first order microscopic transport equations inside the electrode particles with an appropriate Dirichlet boundary condition. }

{\paragraph{A Dirichlet boundary condition for $c_{s,1}$ on $r=R(x)$.}
The Dirichlet condition on $c_{s,1}$ on the electrode particle surfaces, $r=R(x)$, is found by expanding \eqref{nd_7} to first order and rearranging the resulting expression to obtain
\be
c_{s,1}|_{r=R(x)} =\C_1(x,t) \quad \mbox{where} \quad \C_1(x,t)=\frac{\phi_{s,1}(x,t)- \eta_0(x,t)-\phi_0(x,t)}{U'_{eq}( c_{s,0}|_{r=R(x)}) }.  \label{cs1_bc}
\ee
In turn, an expression for $\eta_0$  may be found from the leading order expansion of \eqref{nd_6} 
\be
\eta_0(x,t)=2 \mbox{arcsinh} \left( \frac{G_0(x,t)}{2 \Upsilon \left[ c_{s,0}|_{r=R(x)} (1-c_{s,0}|_{r=R(x)}) c_0(x,t) \right]^{1/2} } \right). \label{eta0}
\ee}

{\paragraph{The leading order electrolyte problem for $c_0$ and $\phi_0$.} In the expression \eqref{cs1_bc} both $\C_1(x,t)$ and $\eta_0$ depend on the leading order solution in the electrolyte. On substituting the expansion \eqref{exp1} into \eqref{nd_1}-\eqref{nd_4} we see that this satisfies the following problem:
\be 
\epsilon_l(x) \mathcal{N}\frac{\partial c_0}{\partial t}+\frac{\partial \Fmz}{\partial x}=0, \quad \Fmz=-{\mathcal B}(x) D(c_0)\frac{\partial c_0}{\partial x}-\Gamma(1-t^+)j_0, \label{lo_liq1}\\
\frac{\partial j_0}{\partial x}=\left\{ \begin{array}{llc}0 & \mbox{for} & -\LL_s<x<0\\
b(x) G_0(x,t)& \mbox{for} & 0<x<1 \end{array} \right. , \label{lo_liq2}\\
j_0=-\mathcal{P}\kappa(c_0) {\mathcal B}(x) \left( \frac{\partial \phi_0}{\partial x}-2\frac{1-t^+}{c_0}\frac{\partial c_0}{\partial x}\right), \label{lo_liq3} \\
 j_0|_{x=-\LL_s} = I(t),  \quad  \Fmz|_{x=-\LL_s} = 0, \quad \phi_0|_{x=-\LL_s} =0, \quad \Fmz|_{x=1}=0, \quad c_0|_{t=0}=1. \label{lo_liq4}
 \ee}

{\paragraph{The first order problem for lithium transport in the electrode.} 
Substituting the expansion \eqref{exp1} into the microscopic transport equations \eqref{nd_8}-(\ref{nd_9}a) and appending the Dirichilet boundary condition \eqref{cs1_bc} leads to the following problems for $c_{s,1}(r,x,t)$
\be
\mathcal{Q} \frac{\partial c_{s,1}}{\partial t}&=&\frac{1}{r^2}\frac{\partial}{\partial r}\left(r^2 \frac{\partial }{\partial r}(D_s(c_{s,0}) c_{s,1})\right) \quad \mbox{in} \quad 0 \leq r \leq R(x),\label{cs1_1} \\
c_{s,1} && \mbox{bounded on} \,\, r=0,  \quad   \left. c_{s,1} \right|_{r=R(x)}=\C_1(x,t), \label{cs1_2}\quad c_{s,1}|_{t=0} =0,  \label{cs1_3} 
\ee
where $\C_1(x,t)$ is defined in \eqref{cs1_bc} and the initial condition comes from expanding the initial conditions \eqref{nd_ic}. By solving this  problem we can find an expression for $ \left. \frac{\dd}{\dd r} (D_s(c_{s,0}) c_{s,1})  \right|_{r=R(x)}$, as a function of $\C_1(x,t)$, that we can use to determine $G_1(x,t)$ using \eqref{G1expr}.}

{\paragraph{Solution of the first order lithium transport problem in the electrode and determing $V_1$.} The problem \eqref{cs1_1}-\eqref{cs1_2} for $c_{s,1}$ is linear.  It can be shown using a Green's function approach that the quantity we seek satisfies an integral equation of the form
\be
\left. \frac{\dd}{\dd r} (D_s(c_{s,0}) c_{s,1})  \right|_{r=R(x)} = \int_0^t \C_1(x,\tau) \G (x,t;\tau) d \tau, \label{green1}
\ee
where the problem  for the `Green's function' $\G (x,t;\tau)$ is derived in Appendix \ref{appb}. In practice deriving this Green's function for a uniformly varying $R(x)$ is extremely costly and arguably more effort that solving the original PET model. However in the notable case of a uniform particle size throughout the half-cell, \ie $R(x) \equiv 1$ the `Green's function'  is independent of $x$ so that in this case
\be
\left. \frac{\dd}{\dd r} (D_s(c_{s,0}) c_{s,1})  \right|_{r=1} = \int_0^t \C_1(x,\tau) \G (t;\tau) d \tau \quad \mbox{for a uniform electrode}.\label{green2}
\ee
Substituting \eqref{green1} into \eqref{G1expr} and the resulting expression back into the solvability condition \eqref{solv_cond} leads to the result
\be
\int_0^1 b(x)  \int_0^t \C_1(x,\tau) \G(x,t; \tau) d \tau dx=0. \label{solv_cond2}
\ee
which provides an avenue to calculate $V(t)$ if we could solve for the Green's function  $\G(x,t; \tau)$.}

{\subsubsection{First order correction to the Single Particle Model} One special case where this result is useful is that of uniform particle size throughout the half cell, \ie $R(x) \equiv 1$ where $\G(x,t;\tau)=\G(t;\tau)$. In this instance the spatial integral can be separated from  the temporal integral in \eqref{solv_cond2} leading to the conclusion that
\be
\int_0^1 b(x) \C_1(x,\tau) dx=0.
\ee
Substituting for $\C_1(x,t)$ from \eqref{cs1_bc}, and for $\phi_{s,1}$ from \eqref{phis1} in the above and rearranging the resulting expression leads to the following formula for the first order correction to the cell voltage:
\be
V_1(t) =\frac{\displaystyle \int_0^1 b(x) \left[\eta_0(x,t) +\phi_0(x,t)- \int_x^1 \frac{j_{s,0}(x',t) }{\Theta\sigma(x')} dx'\right] dx}{\displaystyle \left(\int_0^1 b(x) dx \right) }.
\ee 
Here $\eta_0(x,t)$ is given by \eqref{eta0} and $j_{s,0}(x,t)$ by \eqref{phis1}.}

{\subsubsection{First order correction in the small current limit $\Q \ll 1$} In the case where particle size is non-uniform (\ie $R(x)$ is non constant) we can still make progress in determining the voltage correction if the current is relatively small, corresponding to a small value of $\Q$. Here we seek series solutions for $c_{s,0}(r,x,t)$ and $c_{s,1}(r,x,t)$ in powers of $Q$ to \eqref{lo4}-\eqref{lo6} and \eqref{cs1_1}-\eqref{cs1_3}, respectively. We find that  $c_{s,0}(r,x,t)$ has an expansion in $Q$ of the form
\be
c_{s,0}(r,x,t) &=& \psi(x,t) + \frac{\Q G_0(x,t)}{2 D_s(\psi) R(x)} (R^2(x)-r^2) +\cdots, \\
\mbox{where} \qquad \frac{\dd  \psi}{\dd t} &=& - \frac{3 G_0(x,t)}{R(x)}.
\ee
It follows from \eqref{cs1_1}-\eqref{cs1_3} that
\be
c_{s,1}(r,x,t) &=& \C_1(x,t) - \Q \left( \frac{R^2(x)-r^2}{6 D_s(\psi)} \left( \frac{\dd \C_1}{\dd t}  - \C_1 \frac{D_s'(\psi)}{D_s(\psi)} \frac{\dd  \psi}{\dd t} \right) \right)+\cdots,
\ee
and hence the term $D_s(c_{s,0}) c_{s,1}$ that appears in the right-hand side of \eqref{cs1_1} has the expansion
\bes
D_s(c_{s,0}) c_{s,1}= D_s(\psi(x,t)) \C_1(x,t)+ Q \frac{\dd \C_1}{\dd t} \left( \frac{r^2-R^2(x)}{6} \right) + O(\Q^2).
\ees
Substituting the above into \eqref{G1expr} leads to the follows expression for $G_1(x,t)$:
\be
G_1(x,t)= -\frac{R(x)}{3} \frac{\dd \C_1}{\dd t} + O(\Q).
\ee
Substitution of this expansion int the solvability condition \eqref{solv_cond} and integrating the result with respect to time yields the integral conditon
\be
\int_0^1 b(x) R(x) ( \phi_{s,1}(x,t)-\phi_0(x,t) -\eta_0(x,t) )dx= O(\Q).
\ee
On substituting the expression that we have obtained for $\phi_{s,1}(x,t)$ in \eqref{phis1} into the above and rearranging we obtain the following expression for $V_1(t)$:
\be
V_1(t)=\frac{\displaystyle \int_0^1 b(x) R(x)  \left[\eta_0(x,t) +\phi_0(x,t)- \int_x^1 \frac{j_{s,0}(x',t) }{\Theta\sigma(x')} dx'\right] dx}{\displaystyle \left(\int_0^1 b(x) R(x) dx \right) } +O(\Q).
\ee
}

\section{\label{conc}Conclusions}
{We have shown how the widely-used SPM can be derived directly from the PET model, aka the Newman model, using systematic asymptotic methods as well as how it can be generalised to treat graded electrodes and electrodes with multiple chemistries. We showed that while the results of the SPM model, and its generalisations, give reasonable agreement to the full PET model for relatively small discharge rates the discrepancies become significant at higher rates of discharge. This motivated us to derive a correction term to the SPM model and its generalisation. We demonstrate that the corrected SPM model offers very significantly increased accuracy over the basic SPM model, and its generalisations, to such a degree that it is able to accurately reproduce discharge curves with C-rates up to around 12C in graphite and NMC. Perhaps somewhat surprisingly the corrected SPM model also works reasonably well for LFP even though the asymptotic derivation is not applicable to a material with such a flat discharge curve. However the agreement between the corrected double particle model and the full PET model is not so good for a graded LFP electrode with two different particle sizes.}

{Calculating this correction leads us to what we term the corrected SPM (or a generalisation thereof) which requires that a one-dimensional model for the electrolyte is solved. It is therefore more complex than the basic SPM but is still much cheaper to solve than the full PET model. 
After discretising in space with $N$ mesh points for the electrode particle problem and $2N$ for the electrolyte problem it is found that the total complexity of the corrected SPM is proportional to $N^2$ (meaning that computation times grow like $N^2$) whereas that for the full PET model is proportional to $N^4$. As an example, with $N=50$, we find that a typical solution of the corrected SPM, to obtain a full discharge curve on a standard desktop computer takes approximately 1.5 sec. This compares to a computation time of 170 sec., for the equivalent calculation, using the full PET model, which is two orders of magnitude greater. Furthermore this disparity becomes even more pronounced with increases in $N$.}

So long as the caveats that we have outlined above are respected, the extended versions of the SPM derived here provide an accurate and fast means of modelling the internal {electrochemical} process in modern LIB cells. {The relative ease with which these simulations can be carried out make this type of model an excellent candidate for use in applications where computational costs need to be kept to a minimum such as battery management systems (\eg \cite{plett04}), optimal cell design (\eg \cite{cheng19}), extension of PET to 3-macroscopic- dimensions (\eg \cite{Yi13,Dan16}), and as a tool in parameter estimation studies (\eg \cite{bizeray18,sethurajan19}).}

\paragraph{Acknowledgements}  GR, IK and JF were supported by the Faraday Institution Multi-Scale Modelling (MSM) project Grant number EP/S003053/1. MC acknowledges funding from the University Alliance, Doctoral Training Alliance. The authors are very grateful for the help of Dr. Simon O'Kane in fitting data for the electrode and electrolyte properties.

\appendix
\section{\label{appb} The Green's function for the single particle problem}
Here we write down the problem for the Green's function $\G(r,x,t;\tau)$ for the problem \eqref{cs1_1}-\eqref{cs1_3} that is used to write down the expression for $({\dd}/{\dd r}) (D_s(c_{s,0}) c_{s,1}) |_{r=R(x)}$ in \eqref{green1}. 
We start by seeking a solution $\Gh(r,x,t;\tau)$ that satisfies the problem
\be
\mathcal{Q} \frac{\partial \Gh}{\partial t}&=&\frac{1}{r^2}\frac{\partial}{\partial r}\left(r^2 \frac{\partial }{\partial r}(D_s(c_{s,0}) \Gh)\right) \quad \mbox{in} \quad 0 \leq r \leq R(x),\label{gr_1} \\
\Gh(r,x,t;\tau) && \mbox{bounded on} \ \ r=0,  \quad   \left. \Gh(r,x,t;\tau) \right|_{r=R(x)}=\delta(t-\tau), \label{gr_2}\\
\Gh(r,x,t;\tau)&=&0 \qquad \mbox{for} \quad t \leq \tau.  \label{gr_3} 
\ee
Then by using the superposition principle it is straightforward to show that
\be
c_{s,1}(r,x,t) = \int_0^{\infty} \C_1(x,\tau) \Gh(r,x,t;\tau) d \tau= \int_0^{t} \C_1(x,\tau) \Gh(r,x,t;\tau) d \tau,
\ee
is a solution to the problem \eqref{cs1_1}-\eqref{cs1_3}. It follows that the quantity for which we wish to obtain an expression: $({\dd}/{\dd r}) (D_s(c_{s,0}) c_{s,1}) |_{r=R(x)}$ can be written in the form
\be
\left. \frac{\dd}{\dd r} (D_s(c_{s,0}) c_{s,1})  \right|_{r=R(x)} = \int_0^t \C_1(x,\tau) \G (x,t;\tau) d \tau\\
\mbox{where} \qquad \G (x,t;\tau) =\left. \left[ D_s'(c_{s,0}) \frac{\dd c_{s,0}}{\dd r} \Gh(r,x,t;\tau) +D_s(c_{s,0}) \frac{\dd \Gh}{\dd r}(r,x,t;\tau) \right] \right|_{r=R(x)}.
\ee

\end{document}